\documentclass[a4paper,fleqn,usenatbib]{mnras}

\usepackage{newtxtext,newtxmath}

\usepackage[T1]{fontenc}
\usepackage{ae,aecompl}

\usepackage{graphicx}	
\usepackage{amsmath}	
\usepackage{layouts}
\usepackage{multirow}
\usepackage{xcolor}

\usepackage{color}
\usepackage{hyperref}
\usepackage{supertabular}
\hypersetup{colorlinks, linkcolor=blue}
\usepackage{amsfonts} 
\usepackage{amsmath}
\usepackage{bm}
\usepackage{xcolor}
\usepackage{gensymb}

\def\msun{{\rm M}_{\sun}}
\defcitealias{lagattuta2022}{L22}
\defcitealias{lagattuta19}{L19}
\defcitealias{lagattuta17}{L17}

\title[A BUFFALO view of Abell\,370]{\textit{Beyond the Ultra-deep Frontier Fields And Legacy Observations} (BUFFALO): a high-resolution strong$+$weak-lensing view of Abell\,370}

\author[A. Niemiec et al.\ 2022]{
A. Niemiec,$^{1,2,3}$\thanks{E-mail: anna.niemiec@lpnhe.in2p3.fr}
M. Jauzac,$^{2,3,4,5}$
D. Eckert,$^{6}$
D. Lagattuta,$^{2,3}$
K. Sharon,$^{7}$
A. M. Koekemoer,$^{8}$ 
\newauthor
K. Umetsu,$^{9}$
A. Acebron,$^{10,11}$
J. M. Diego,$^{12}$ 
D. Harvey,$^{13,14}$
E. Jullo,$^{15}$
V. Kokorev,$^{16,17}$
\newauthor
M. Limousin,$^{18}$
G. Mahler,$^{2,3}$  
P. Natarajan,$^{18,19}$ 
M. Nonino,$^{20}$ 
C. Steinhardt,$^{21}$
S.-I. Tam,$^{9}$ 
\newauthor
A. Zitrin$^{22}$  
\\
\\
\\
$^{1}$LPNHE, CNRS/IN2P3, Sorbonne Universit\'e, Universit\'e Paris-Cit\'e, Laboratoire de Physique Nucl\'eaire et de Hautes \'Energies, 75005 Paris, France\\
$^{2}$Centre for Extragalactic Astronomy, Durham University, South Road, Durham DH1 3LE, UK\\
$^{3}$Institute for Computational Cosmology, Durham University, South Road, Durham DH1 3LE, UK\\
$^{4}$Astrophysics Research Centre, University of KwaZulu-Natal, Westville Campus, Durban 4041, South Africa \\
$^{5}$School of Mathematics, Statistics \& Computer Science, University of KwaZulu-Natal, Westville Campus, Durban 4041, South Africa\\
$^{6}$Department of Astronomy, University of Geneva, Ch. d'Ecogia 16, 1290 Versoix, Switzerland \\
$^{7}$Department of Astronomy, University of Michigan, 1085 South University Ave, Ann Arbor, MI 48109, USA\\
$^{8}$Space Telescope Science Institute, 3700 San Martin Dr., Baltimore, MD 21218, USA\\
$^{9}$Academia Sinica Institute of Astronomy and Astrophysics (ASIAA), No.~1, Section~4, Roosevelt Road, Taipei 10617, Taiwan\\
$^{10}$Dipartimento di Fisica, Universit\`a  degli Studi di Milano, via Celoria 16, I-20133 Milano, Italy \\
$^{11}$INAF - IASF Milano, via A. Corti 12, I-20133 Milano, Italy \\
$^{12}$Instituto de F\'isica de Cantabria (CSIC-UC). Avda. Los Castros s/n. 39005 Santander, Spain \\
$^{13}$Lorentz Institute, Leiden University, Niels Bohrweg 2, Leiden, NL-2333 CA, The Netherlands \\
$^{14}$Laboratoire d'Astrophysique, Ecole Polytechnique F\'ed\'erale de Lausanne (EPFL), Observatoire de Sauverny, CH-1290 Versoix, Switzerland \\
$^{15}$Aix Marseille Univ., CNRS, CNES, LAM, Marseille, France  \\
$^{16}$Cosmic Dawn Center (DAWN), Jagtvej 128, DK2200 Copenhagen N, Denmark \\
$^{17}$Niels Bohr Institute, University of Copenhagen, Blegdamsvej 17, DK2100 Copenhagen \O, Denmark \\
$^{18}$Department of Astronomy, Yale University, 52 Hillhouse Avenue, New Haven, CT 06511, USA \\
$^{19}$Department of Physics, Yale University, P.O. Box 208121, New Haven, CT 06520, USA \\
$^{20}$INAF-Trieste Astronomical Observatory , Trieste, Via Bazzoni 2, Italy \\ 
$^{21}$Dark Cosmology Centre, Niels Bohr Institute, Juliane Maries Vej 30, 2100 K\o benhavn, Denmark \\
$^{22}$Physics Department, Ben-Gurion University of the Negev, P.O. Box 653, Be'er-Sheva 84105, Israel 
}

\date{Accepted XXX. Received YYY; in original form ZZZ}

\pubyear{2022}

\begin{document}
\label{firstpage}
\pagerange{\pageref{firstpage}--\pageref{lastpage}}
\maketitle

\begin{abstract}
The \emph{HST} treasury program BUFFALO provides extended wide-field imaging of the six \emph{Hubble Frontier Fields} galaxy clusters. Here we present the combined strong and weak-lensing analysis of Abell\,370, a massive cluster at $z=0.375$. From the reconstructed total projected mass distribution in the $6\arcmin \times 6\arcmin$ BUFFALO field-of-view, we obtain the distribution of massive substructures outside the cluster core and report the presence of a total of seven candidates, each with mass $\sim 5\times 10^{13}M_{\odot}$. Combining the total mass distribution derived from lensing with multi-wavelength data, we evaluate the physical significance of each candidate substructure, and conclude that 5 out of the 7 substructure candidates seem reliable, and that the mass distribution in Abell\,370 is extended along the North-West and South-East directions. While this finding is in general agreement with previous studies, our detailed spatial reconstruction provides new insights into the complex mass distribution at large cluster-centric radius. We explore the impact of the extended mass reconstruction on the model of the cluster core and in particular, we attempt to physically explain the presence of an important external shear component, necessary to obtain a low root-mean-square separation between the model-predicted and observed positions of the multiple images in the cluster core. The substructures can only account for up to half the amplitude of the external shear, suggesting that more effort is needed to fully replace it by more physically motivated mass components. We provide public access to all the lensing data used as well as the different lens models.
\end{abstract}




\section{Introduction}
As they are fairly easy to detect thanks to the multiple available observational tracers (galaxies, X-ray gas), galaxy clusters are an important probe of the formation history of structures in the Universe. Indeed, they are among the most massive gravitationally bound objects, and therefore represent the latest stage of the hierarchical structure formation, meaning that they have formed over time through the accretion of many smaller objects, i.e., galaxies or  galaxy groups and the dark matter haloes in which they live. In particular, very massive and complex clusters bear the dynamical trace of  their long accretion history, for instance through their multi-modal structure \citep[e.g.,][]{clowe06, merten11, limousin12, finner2021, cho2022, pascale2022, monteiro-oliveira2022} and the number and mass of their substructures \citep[cf.][]{jauzac16b}. Clusters are also excellent laboratories for probing the physical mechanisms governing their evolution. Their extreme densities make them an ideal location to study the nature of dark matter.
 
However, an essential ingredient needed to use clusters as probes of these complex processes is the detailed knowledge of their overall structure and total matter distribution; this needs to be measured first. To obtain an inventory of the baryonic component of clusters, multi-wavelength observations are combined, from the stars in cluster member galaxies and intracluster light (ICL) traced by optical emission from their stellar populations; to the hot gas emitted in the X-rays or backlit by the cosmic microwave background through the Sunyaev-Zeldovich effect. The dark matter component is more elusive and can, for now, only be detected through indirect measurements. One method, and a particularly powerful established one, is gravitational lensing, which probes the gravitational potential of a massive structure, from the bending of light rays that pass through it. Cluster lensing \citep[see][for a review]{kneib&natarajan2011,umetsu2020} can manifest itself in two different regimes, depending on the local matter density. In very dense inner regions, such as the core of clusters, light rays are strongly deflected, and images of background galaxies are distorted into gravitational arcs, or can even appear multiple times, characterizing what is refereed to as the strong lensing regime. In less dense regions of clusters, where light rays are only slightly bent, images of background galaxies are only weakly distorted. As this lensing induced shear is much smaller than the intrinsic ellipticity of a galaxy (and is at the percent level), it is impossible to distinguish both components in the measured ellipticity of individual sources. This so-called weak lensing regime in turn requires a statistical approach to overcome this \emph{shape noise} coming from the galaxy intrinsic ellipticities, to derive the cluster mass distribution.
Cluster lensing, both in the strong and weak regimes, has shown to be a key tool in many studies, from probing the nature of dark matter \citep[e.g.,][]{natarajan02b,natarajan17,bradac08a,harvey15,harvey16,harvey17,harvey2019,massey15,massey18,jauzac16b,jauzac18b, meneghetti2020, vega-ferrero2021, andrade2022, bhattacharyya2022,  limousin2022}, to constraining cluster physics \citep[e.g.,][]{kneib03,natarajan02a,clowe04,bradac06,merten11,diego15a,eckert15,grillo15,jauzac12,jauzac15a,mahler18a,mahler2020,sharon15,sharon2020,bergamini2021, moura2021, ebeling2021, chadayammuri2022, fox2022} or galaxy evolution \citep[e.g.][]{natarajan98,limousin07a,natarajan09,li2015, sifon2015, niemiec2017, sifon2018}. 
In addition to probing the internal structure of galaxy clusters, lensing also offers a new window for observing the distant Universe. As lensing magnifies the images of background galaxies, and cluster lenses therefore act as cosmic telescopes, making it possible to probe star formation within galaxies at cosmic noon \citep[e.g.][]{rigby2017, rigby2018, rigby2021, johnson2017, chisholm2019, man2021, vanzella2022, ditrani2022, furtak2022}, and to observe and study very distant galaxies by bringing them into view \citep[e.g.][]{daloisio2014,atek15,atek18,alavi2016,bouwens2017,ishigaki18,kawamata18,salmon2018,salmon2020, furtak2021, strait2021, laporte2021, bouwens2022a, bouwens2022b, sun2022, yang2022}.

In this larger context the executed \emph{Hubble Frontier Fields} programme \citep[HFF,][]{lotz17},  provided the deepest images of galaxy clusters ever obtained, using the \emph{Hubble Space Telescope} (HST). HFF targeted six particularly massive clusters, and devoted a total of HST 840 orbits to the 6 clusters, imaging from the UV to the near-infrared (7 filters), covering the core and a parallel field located 4\,arcminutes from the core. HFF observations have led and continue to lead to a vast number of studies, covering both cluster-related and high redshift science \citep[e.g.][]{richard14, johnson14, grillo15, kawamata16, meneghetti17, bouwens17b, ishigaki18, montes2019, richard2021}.
  
The BUFFALO \citep[\emph{Beyond the Ultra-deep Frontier Fields and Legacy Observations}, GO-15117, P.I.s: Steinhardt \& Jauzac,][]{steinhardt2020} survey was designed to build upon the success of the HFF campaign. It is a large \textit{HST} programme that extends the coverage of the six HFF clusters with the \textit{Advanced Camera for Survey} (ACS), in the F814W and F606W pass-bands, as well as with the \textit{Wide Field Camera 3} (WFC3), in the F105W, F125W and F160W pass-bands. The main scientific goals of the survey are twofold: (i) to study the foreground clusters and improve their mass modelling, by adding deep and high-resolution weak-lensing observations to the already existing very deep strong-lensing constraints, and (ii) to extend the observed area available for magnified high redshift background source studies. These overarching objectives encompass many different science cases, such as probing the physical processes in galaxy clusters or the properties of dark matter haloes, mapping the substructure distribution or the intra-cluster light, studying galaxy evolution in the cluster environment, measuring the high redshift UV luminosity function or the star formation rate-stellar mass relation, studying high redshift quiescent galaxies, etc. Our focus in the work presented here is to report on the major gains and resulting improvement of the modeling of the cluster total mass distribution using a combination of strong- and weak-lensing constraints to model the core and the outskirts simultaneously.
 
In this paper, we focus on the first fully observed cluster from the BUFFALO survey, Abell\,370. Abell\,370 is a very massive cluster with $M_{200} = (2.2 \pm 0.3) \times 10^{15} h^{-1}\msun$, and located at a redshift $z = 0.375$. Its mass and total matter distribution have been modelled by multiple teams previously, using strong \citep{richard10a, lagattuta17, lagattuta19, ghosh2021} or weak-lensing constraints \citep{medezinski10, strait2018, umetsu2022}. 

\cite{lagattuta19}  \citepalias[hereafter][]{lagattuta19} presented a strong lensing model of the cluster core, using a combination of the HFF observations and integral-field spectroscopy obtained with the Multi Unit Spectroscopic Explorer (MUSE) at the Very Large Telescope (VLT). Interestingly, their final model integrates an external shear component. Such components are often introduced in strong-lensing analyses, and produce a uniform shear on the position and/or shape of the constraints in order to improve the goodness-of-fit of the model. These components can be difficult to justify physically: they are often not directly linked to identified mass components, and in addition these components are generally uniform over the whole modeled field, which is not easily reproducible with mass distribution. This can come from the fact that all model decompositions are only approximations of the underlying mass distribution. The physical interpretation of the external shear components should therefore be treated with care, as they can be an approximation of the impact of some (sub)structures, but can also be the result of other approximations in the modeling, such as the limited choice in terms of potential shapes. In \citetalias{lagattuta19}, to physically motivate the presence of this shear, required in their model to properly reproduce the observed position of the multiple image sources, they explore the impact of line-of-sight structures on the mass model, but could not account for the full external shear term with that component.

Another physical explanation for this term could be the presence of some massive structures in the neighborhood of the cluster \citep[][]{acebron17}, or the presence of some unmapped massive substructures within the cluster itself. The importance of weak-lensing data in the cluster outskirts is tantamount in this situation, as it allows us to map the projected total mass distribution outside the cluster core, and possibly detect the substructures that could generate this needed shear in the cluster core. Such a case was presented for instance in \citet{mahler18a} for another HFF cluster, Abell~2744. They found that including the substructures detected in the cluster's outskirts in \citet{jauzac16b} to their strong-lensing model, could replace an external shear component, and improve the model significantly, compared to a no-substructure no-external shear version. Similarly, \citet{kawamata16} present a mass reconstruction for another HFF cluster MACS\,0717 that also required an external shear component, but in this case they argued that the physical origin for this component was probably a structure located along the line-of-sight.

It is critically important to understand the origin of the external shear in these mass model configurations, as this component is uniform in strength and in direction over the entire modeled field, making it very challenging to interpret physically, especially when  modeling large fields as in the present case. In this regard, it is first crucial to model the cluster mass distribution at all scales, from the extremely dense cluster core to the potentially irregular outskirts. For this we use the publicly available lens modeling code \textsc{Lenstool} \citep{jullo07,jauzac12,jullo14}, and in particular the new version \emph{hybrid}-\textsc{Lenstool} that we developed to take these complexities into account \citet{niemiec2020}. 
In general, self-consistently modelling the mass distribution in clusters at all scales is crucial to limit environment-induced biases on the models describing the cluster core; to derive more accurate and precise magnification estimates in the outskirts; and better constrain galaxy and cluster evolution, etc. The BUFFALO data set, combined with suitable modelling techniques, presents a unique opportunity for such studies.

The goal of this paper is to measure the total mass distribution of Abell\,370 in the entire BUFFALO field, which covers approximately $6' \times 6'$ (or $\sim 1.9 \times 1.9 \,\mathrm{Mpc}^2$ at the cluster redshift), powerfully combining strong- and weak-lensing constraints. The model includes the core of the cluster and the outskirts, with two main objectives: (i) detect all possible substructures, to better understand the structure and dynamical evolutionary history of this cluster, and (ii) test how the detected mass distribution in the outskirts impacts the model in the core, and in particular, if it allows us to reduce or completely remove the external shear component by explicitly accounting for it with detected substructures.  To achieve these goals, we test different modelling methods to verify if, and how, our modelling assumptions impact the results. We also compare the projected total mass distribution with X-ray observations from \textit{XMM-Newton}, as well as galaxy dynamics, to corroborate the physical existence of the lensing-detected substructures.

This paper is organised as follows: in Sect.~\ref{sec:obs} we present the \emph{HFF} and \emph{BUFFALO} observations of Abell\,370, and describe the construction of the strong- and weak-lensing constraints catalogues in Sect.~\ref{sec:slwl}. We summarise the modeling methods in Sect.~\ref{sec:model}, present the total mass distribution obtained with our two main modelling methods in Sect.~\ref{sec:sequential_fit} and ~\ref{sec:joint_fit}, and complement this in Sect.~\ref{sec:baryons} with the baryonic mass distribution derived from optical and X-ray observations. We then evaluate the impact on the model of the substructures detected in the mass reconstructions in Sect.~\ref{sec:subs_impact}. Finally, given the different mass models and the mutli-wavelength analysis, we discuss in Sect.~\ref{sec:discussion} the physical reality of the candidate substructures, and whether their presence suffice to account for the external shear component in the models.  
Throughout this paper, we use a standard flat $\Lambda$CDM cosmology with $\Omega_{\rm{m}} = 0.27$ and $h_0 = 0.7$. In this cosmology, 1"=5.2\,kpc at the cluster redshift of z=0.375. All magnitudes are quoted in the AB system \citep{oke74}.

\section{Observations}
\label{sec:obs}

\subsection{\emph{Hubble Space Telescope} observations}

    \subsubsection{Hubble Frontier Fields}
Abell\,370 has been extensively observed with \textit{HST} over the past two decades. 
Observations were initially taken with the \emph{Advanced Camera for Survey} (ACS) in the F814W pass-band (GO-11507; P.I.: Noll), then in the near-infrared with the \emph{Wide Field Camera 3} (WFC3) in the F140W pass-band (GO-11108; P.I.: Hu). It was then observed with both ACS and WFC3 with GO-11591 (PI: Kneib), GO-13459 (PI: Treu), GO-14038 (PI: Lotz), and GO-14216 (PI: Krishner).

It is the HFF \citep[][]{lotz17} observing campaign that has provided the deepest observations of its core and a parallel field 4\arcmin\ away from its central brightest cluster galaxy (BCG) with \emph{HST}. The HFF campaign observed Abell\,370 over 140 orbits during Cycle 25, in 7 pass-bands from the UV to the near-infrared (F435W, F606W, F814W, F105W, F125W, F140W, F160W). The details of the HFF observing campaign are presented in \cite{lotz17}.

\subsubsection{{Beyond the Ultra-deep Frontier Fields And Legacy Observations}}

\begin{figure*}
\begin{center}
\includegraphics[width=\textwidth,angle=0.0]{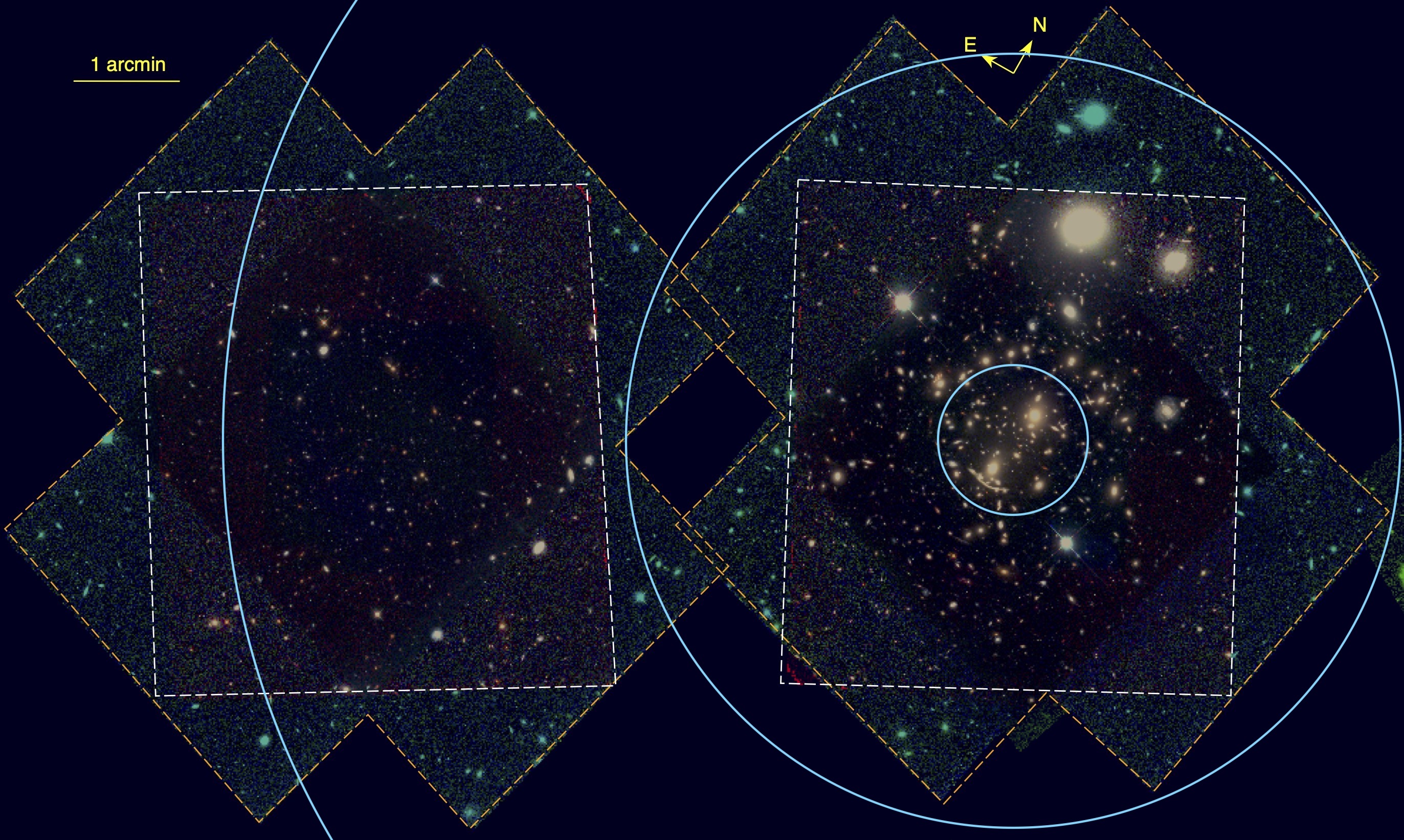}
\caption{Full BUFFALO colour-composite mosaic for Abell\,370, in the F160W, F814W and F606W filters. The dashed orange (white) footprints show the approximate field-of-view of the ACS (WFC3) observations. The deepest observed squares in the middle of each field correspond to the HFF observations. The main field, on which we focus the anlysis presented here, is shown on the right-hand side of the image, and the parallel field is shown on the left-hand side. Considering that for Abell\,370 the virial radius is $R_{200} = 2.3$\,Mpc \citep{lagattuta2022}, we show as blue circles the approximate positions of $0.1R_{200}$, $0.5R_{200}$ and $R_{200}$. 
}
\label{fig:buffalo_fov}
\end{center}
\end{figure*}

The HFF was used as the baseline for the \emph{Beyond the Ultra-deep Frontier Fields And Legacy Observations} \citep[BUFFALO; GO-15117; PIs: Steinhardt \& Jauzac,][]{steinhardt2020}. BUFFALO extends the spatial coverage around both the core and the HFF parallel field, providing an almost continuous field of view of $\sim$13$\times$8\,arcmin$^{2}$ in the ACS/F814W and F606W pass-bands, and $\sim$10$\times$5\,arcmin$^{2}$ in the WFC3/F105W, F125W and F160W pass-bands, with a gap of $\sim$2$\times$5\,arcmin$^{2}$ between the core and the parallel field. The first epoch of Abell\,370 observations was taken between 2018 July 21 and August 21, and the second epoch between 2018 December 19 and 2019 January 31, for 5180\,s, 9428\,s, 5647\,s, 6447\,s and 6447\,s with ACS/F606W, ACS/F814W, WFC3/F105W, WFC3/F125W and WFC3/F160W pass-bands, respectively. 
The mosaics that we use here include all the new BUFFALO data \citep{steinhardt2020} as well as the archival data such as HFF \citep{lotz17} and were produced following the approaches first described by \citet{koekemoer2011}, extending significantly beyond the standard pipeline archive products.
The full BUFFALO mosaic in the WFC3/F160W, ACS/F814W and ACS/F606W filters, as well as the field-of-view of the ACS and WFC3 observations are shown in Fig.~\ref{fig:buffalo_fov}.

\subsection{Subaru/Suprime-Cam observations}

To complement the deep BUFFALO imaging covering the central region of the cluster, we also use in our analysis a weak-lensing and photometric catalogue obtained from deep Subaru/Suprime-Cam in the $BR_{\rm{C}}z'$-bands, and the instantaneous field of view of
Suprime-Cam is 34 arcmin x 27 arcmin. The details of the observations and data reduction are described in \citet{umetsu2022}.

\subsection{X-ray observations}
\label{sec:xray_data}

\subsubsection{Data reduction}

Abell\,370 was observed by \emph{XMM-Newton} on 2017 January 22 for a total of 133 ks (Observation ID 0782150101). We reduced the observation using XMMSAS v19.1 and the analysis pipeline developed in the framework of the \emph{XMM-Newton} cluster outskirts project \citep[X-COP,][]{xcop}. We ran the standard event screening chains to extract calibrated event files for the three detectors of the European photon imaging camera (EPIC): MOS1, MOS2, and PN. For each camera, we extracted light curves of the observations and filtered out time periods affected by soft proton flares using the XMMSAS tasks \texttt{mos-filter} and \texttt{pn-filter}. The remaining clean exposure time amounts to 57.5 ks (MOS1), 59.3 ks (MOS2), and 45.1 ks (PN). 

From the clean event files, we extracted images in 5 energy bands (0.4-0.7, 0.7-1.2, 1.2-2.0, 2.0-4.0, and 4.0-7.0 keV) and used the XMMSAS tasks \texttt{mos-spectra} and \texttt{pn-spectra} to model the spatial distribution and intensity of the quiescent particle background in each band. This is achieved by measuring the high-energy count rate in the unexposed corners of the three cameras and rescaling filter-wheel-closed data available in the calibration database to match the observed count rate. Background maps are then extracted using the same binning as the actual images from the rescaled filter-wheel-closed event files. Exposure maps were created using the \texttt{eexpmap} executable, which computes the local effective exposure time accounting for vignetting, chip gaps and dead pixels. We then stacked the maps from the three individual cameras and combined their exposure maps weighted by their corresponding effective area. The resulting maps combine all the available clean EPIC data. Finally, we used \texttt{asmooth} \citep{ebeling06} to create adaptively-smoothed, vignetting-corrected, and background-subtracted surface brightness maps. For more details on the data analysis procedure see \citet{ghirardini19}.

\begin{figure}
\begin{center}
\includegraphics[width=0.5\textwidth,angle=0.0]{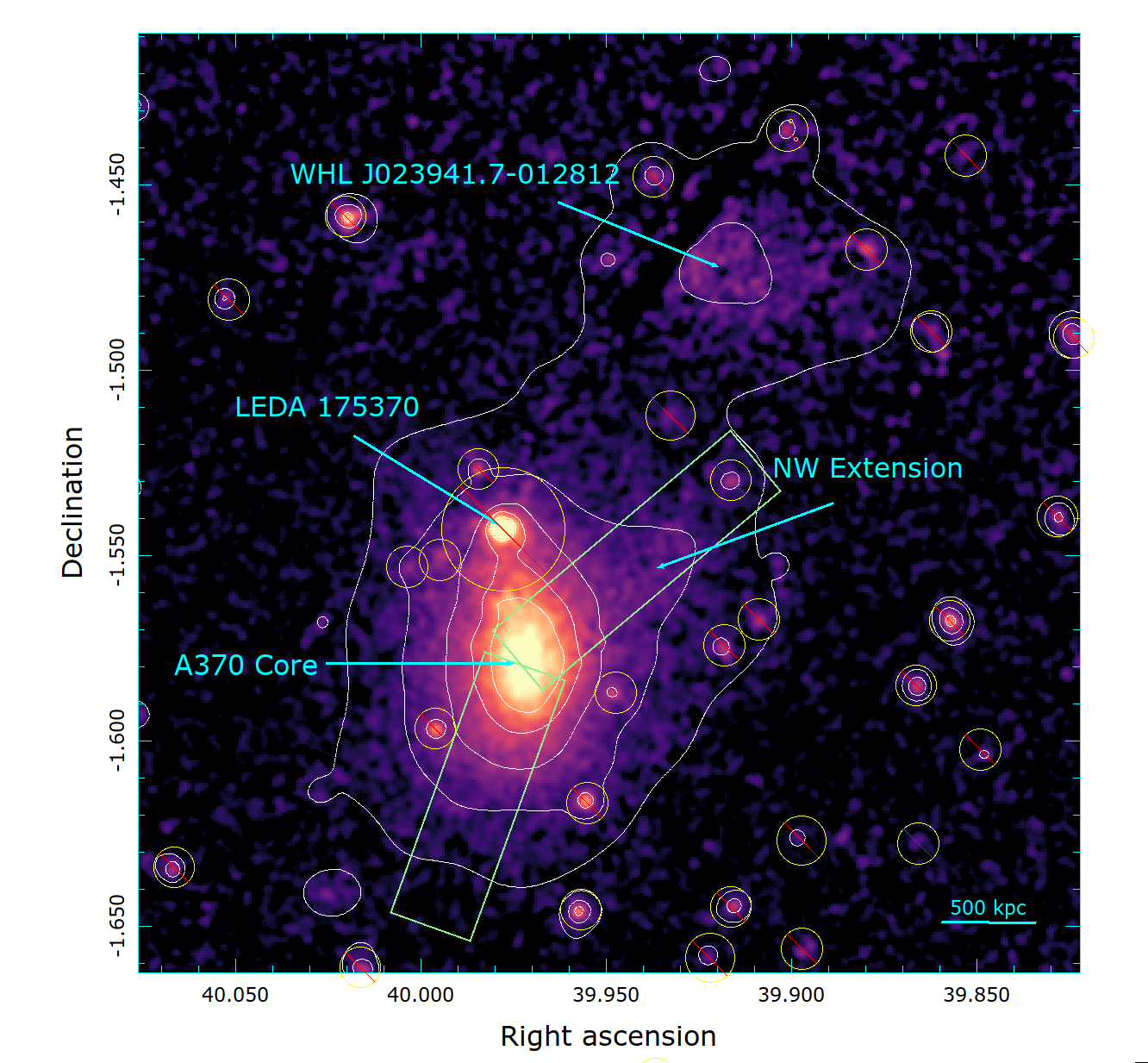}
\caption{Background-subtracted and vignetting-corrected \emph{XMM-Newton} surface brightness map in the [0.5-2] keV band smoothed with a Gaussian of $\sigma=3$ pixel (7.5 arcsec). The white contours are the X-ray isocontours extracted from our adaptively smoothed map (see text). The barred circles show the positions of point sources and their corresponding exclusion areas. The positions of several striking features are highlighted with the cyan arrows. The green boxes show the regions used for the extraction of the surface brightness profiles.
}
\label{fig:xray_map}
\end{center}
\end{figure}

In Fig.~\ref{fig:xray_map} we show the background-subtracted and vignetting-corrected \emph{XMM-Newton} map of Abell\,370 in the [0.5-2] keV band. The cluster appears highly elliptical with an elongation along the North-South axis. An important complication for the analysis is the presence of the bright foreground galaxy LEDA 175370 ($z=0.045$) located $\sim2$ arcmin North of the cluster core, which is associated with a bright, spatially extended X-ray source as highlighted on Fig. \ref{fig:xray_map}. On top of that, we clearly detect a clump of diffuse X-ray emission $\sim7$ arcmin NW of the core of Abell\,370 (see the discussion in Sect. \ref{sec:xray_north}). In the cluster itself, we observe an extension of low surface brightness X-ray emission extending in the NW direction from the cluster core, which is highlighted as well on Fig. \ref{fig:xray_map}. Finally, point sources detected on the [0.5-2] keV maps using the \texttt{ewavelet} task are shown on the same figure and masked for the remainder of the analysis. 

\subsubsection{Thermodynamic maps}

We used the available images in 5 energy bands together with their corresponding exposure and background maps to extract thermodynamic maps of the cluster (temperature, emission measure, pressure, and entropy). To this aim, we used the advance plasma emission code \citep[APEC,][]{smith01xray} folded with the \emph{XMM-Newton} response files to create spectral templates integrated in the 5 bands of interest as a function of the plasma temperature \citep[][]{jauzac16b}. The metallicity of the gas was fixed to $0.3Z_\odot$ and the APEC model was absorbed by photo-electric absorption to model the absorption of photons along the line of sight by the Galactic column density, which was fixed to the value of $2.89\times10^{20}$ cm$^{-2}$, estimated from the 21-cm map in the region surrounding Abell\,370 \citep[][]{HI4PI}. 

Around each pixel for which the local 0.5-2 keV surface brightness exceeds the background surface brightness by more than $3\sigma$, we accumulated total counts in the full band (0.4-7.0 keV) within a circular region surrounding the pixel until the total normal of counts reaches a threshold of 200 counts. We then measured the surface brightness in each of the 5 bands after having masked the relevant point sources, and we fit the spectral energy distribution with the APEC templates by minimizing the C-statistic. The adaptive nature of the binning scheme naturally implies that neighbouring points are correlated, with a correlation length that is equal to the radius of the circular region defined around each bin. In the case of Abell\,370, the correlation length goes from 10 arcsec in the cluster core to $\sim 50$ arcsec in the cluster outskirts. 

The diffuse emission from the LEDA 175370 galaxy likely extends over several arcmin and its soft spectrum biases the measured temperatures low in the neighbouring regions. In our temperature map, we mask a circle of $1$ arcmin radius around the galaxy; however, it is still likely that the cluster temperatures in the Northern part of the cluster are somewhat underestimated.

\subsubsection{The Northern clump}
\label{sec:xray_north}
A striking feature detected in our X-ray map is the presence of an extended clump of diffuse low surface brightness emission 1.5 Mpc in projection NW from the core of Abell\,370. The diffuse X-ray source is associated with the photometric optical cluster WHL J023941.7-012812 \citep{wen15} at a redshift of 0.325. Although the indicated redshift places the system relatively far in projection from Abell\,370, given the uncertainties associated with photometric redshift estimations we cannot exclude that the system is part of the same superstructure as the main cluster. The Northern clump appears as well on the Subaru WL map of \citet{umetsu2022}, indicating it is a massive structure. 

We extracted the \emph{XMM-Newton} spectrum of the Northern clump within a circle of 2 arcmin radius and fitted it with an APEC model at a redshift of 0.325. The best-fit APEC model returns a temperature of $2.1\pm0.2$ keV, which corresponds to a mass of $M_{500}=(1.2\pm0.3)\times10^{14} M_\odot$ using the weak-lensing-calibrated mass-temperature relation of \citet{umetsu20b}  derived for the XXL survey sample. We discuss the impact of this structure on the value of the external shear in the core of Abell\,370 in Sect. \ref{sec:external_shear}.

\subsection{VLT/MUSE observations}
\label{sec:muse_data}

MUSE is an integral field spectrograph installed on the VLT, covering optical wavelengths in the range 465-930\,nm, and a relatively large $1\arcmin \times 1\arcmin$ field-of-view.  \citet{lagattuta17} (hereafter \citetalias{lagattuta17}) used an initial Guaranteed Time Observing (GTO) programme, 094.A-0115(A) (P.I.: Richard) focused on the core of the cluster. They then combined it with the observations from the programme 086.A-0710(A) (P.I.: Bauer), expanding the initial GTO observations. The final MUSE mosaic covers an area of 2$\times$2\,arcmin$^{2}$. The final spectroscopic redshift catalogue published in \citetalias{lagattuta19}, including multiply or singly lensed background sources, cluster members and foreground objects, contains 584 objects, but only 506 unique systems when accounting for multiply imaged background galaxies.

\citet{lagattuta2022} (hereafter \citetalias{lagattuta2022}) further expanded the Abell\,370 MUSE footprint by observing 10 additional pointings surrounding the original central mosaic (PID: 0102.A-0533(A), P.I.s F.~Bauer and D.~Lagattuta). Though shallower than the data in the core region (having only 1-hr exposure times compared to the 2-8 hr depths in the centre), these pointings represent a $\sim 250\%$ increase in coverage area, and the resulting redshift catalogue generated from this region provides an additional 649  unique redshifts, including 180 cluster members and 109 galaxies in the distant universe ($z > 3$).

\subsection{Ancillary spectroscopic and photometric redshift catalogues}

In order to calibrate the different galaxy selections necessary for the analysis (background galaxies for the weak lensing analysis, cluster member galaxies), we use the galaxies in the BUFFALO field-of-view that have measured redshifts. We combine spectroscopic redshifts coming from different sources:
\begin{itemize}
    \item the MUSE catalogues from  \citetalias{lagattuta19, lagattuta2022} described above;
    \item the Grism Lens-Amplified Survey from Space \citep[GLASS][]{schmidt14,treu15} observations. GLASS is a large spectroscopic programme that targeted among others the HFF clusters with the near-IR G102 and G141 grisms onboard \emph{HST}/WFC3. GLASS observations of Abell\,370 consist of 21859\,s of G102 data and 8326\,s of G141 data (roughly 8 and 3 \emph{HST} orbits, respectively). The final GLASS redshift catalogue contains 511 entries for Abell\,370, though we only use 112 entries in the combined spectroscopic catalogue in this work, due to the large overlap between the GLASS and MUSE footprints. 
    \item And 210 galaxies that were in neither of the previous catalogues but had a redshift on the NASA/IPAC Extragalactic Database\footnote{\url{https://ned.ipac.caltech.edu/}} (NED). 
\end{itemize}
In addition to spectroscopy, we also use photometric redshifts to calibrate the background galaxy selection for the weak lensing catalogue. The photometric redshifts were computed using the BPZ algorithm \citep[][]{benitez2000, benitez04, coe06}.

\section{Gravitational lensing data products}
\label{sec:slwl}
In this section, we present the different data products used in the lens models of Abell\,370: the strongly lensed multiple images, the weakly lensed background sources and the cluster member galaxies.

\subsection{Strong-lensing constraints}
\label{sec:sl_constraints}

\citetalias{lagattuta17} and \citetalias{lagattuta19} presented a strong-lensing analysis of Abell\,370 using the combination of HFF imaging and extended VLT/MUSE observations.  
\citetalias{lagattuta17} identified 22 multiple image systems, only 4 of which had all multiple images contained within the MUSE GTO field of view. \citetalias{lagattuta19} completed this picture by confirming all the counter-image candidates. Adding to that, they spectroscopically identified 18 new multiple image systems in the wider MUSE mosaic, within the redshift range $2.9<z<6.3$, bringing the total of spectroscopically confirmed strong-lensing constraints to 39 multiple image systems (that is 103 images). In addition, they identify and incorporate as model constraints 23 photometric multiple images (from 7 background sources). We note that \citetalias{lagattuta19} performed multiple modeling runs using different sets of strong lensing constraints. In addition to the secure spectroscopically confirmed systems (called \emph{gold}-class systems), they considered less reliable classes of constraints, for instance systems without a secured redshift measurement, subdivided into the \emph{silver-}, \emph{bronze-} and \emph{copper}-class systems, as defined during the HFF public lens modeling challenge. \emph{Silver-}class systems were considered very reliable by the different modeling teams participating in the challenge, but lacked a definitive spectroscopic confirmation, while \emph{bronze} systems not only lacked this information, but were also classified as less secure by the modeling teams. Finally, \emph{copper-}class systems were not considered in the original HFF modeling challenge, but were still believed to be true multiple-image systems in \citetalias{lagattuta19}.
For further details on the different sets of constraints, we refer the reader to \citetalias{lagattuta19}.

As part of the BUFFALO collaboration, these latest constraints were combined with the previous literature \citep{lagattuta17, kawamata18, diego2018}, to produce a unified and updated catalogue of strong lensing constraints. These strong lensing systems were then revoted by different members of the BUFFALO \emph{Mass Modeling Working Group}, following a similar procedure as in the HFF challenge: each modeling team attributed a grade between 1 and 4 to each constraint (1=good; 2=less certain; 3=probably wrong and 4=would not use it). Images with an average vote better than 1.5 were then classified as \emph{gold} or \emph{silver}, depending on the availability of a spectroscopic confirmation. Remaining images with a lower vote were classified as  \emph{bronze}. A new category was also added, to account for the specifics of this data set: the \emph{quartz} constraints. They are systems identified directly in the MUSE observations in \citetalias{lagattuta19}, and which do not have an obvious counterpart in the \textit{HST} images. In terms of reliability, we classify these constraints as being between the \emph{gold} and the \emph{silver}, as they are fairly secure given their MUSE detection and redshift, but lack the high-precision \textit{HST} coordinates. The final BUFFALO sample contains 32 gold, 6 quartz, 8 silver and 12 bronze systems, which corresponds to 98, 18, 21 and 39 images respectively. The position of each image in the catalogue has been carefully re-examined by a few members of the collaboration to ensure that the chosen positions match between each image in a system.  

In this analysis, we use only the most secure (\emph{gold}) class, as our goal is not to re-create a complex strong-lensing model in the core of the cluster from scratch, but to study the distribution of substructures and their impact on the overall modeling. The 98 \emph{gold-}class multiple images span a large redshift range ($0.73 < z < 6.29$) and ensure a high density of constraints to model the cluster core, with an average of 22\,arcmin$^{-2}$. However, the whole set of strong-lensing constraints (\emph{gold, quartz, silver, bronze}) is publicly released together with this paper.


\begin{figure}
\begin{center}
\hspace{-5mm}\includegraphics[width=0.5\textwidth,angle=0.0]{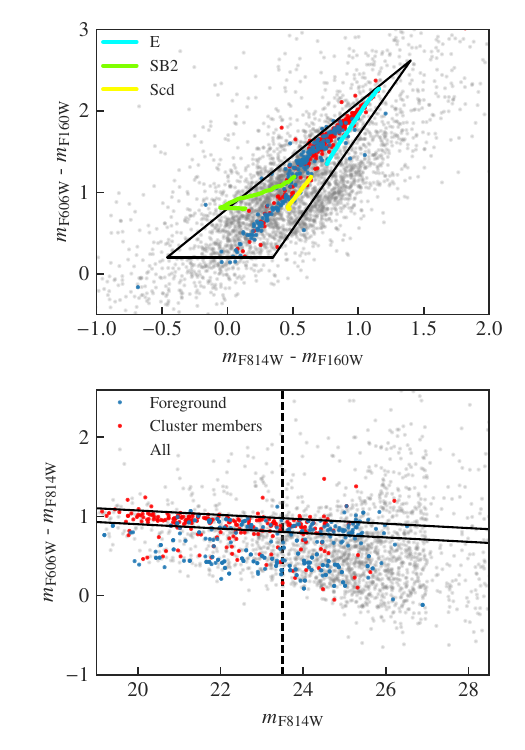}
\caption{\emph{Top:} Colour-colour diagram ($m_{\mathrm{F814W}}$-$m_{\mathrm{F160W}}$) \emph{vs} ($m_{\mathrm{F606W}}$-$m_{\mathrm{F160W}}$) for objects within the BUFFALO ACS and WFC3 overlap field of view for Abell\,370. Grey dots represent all galaxies with both ACS and WFC3 imaging. Unlensed galaxies diluting the shear signal are marked by different colours: galaxies identified as foreground galaxies with either photometric or spectroscopic redshifts $z < 0.335$ (blue); galaxies classified as cluster members due to their spectroscopic or photometric redshifts $0.335 < z < 0.415$ (red). The solid black lines delineate the colour-colour-cut defined for this work to mitigate shear dilution by unlensed galaxies. Theoretical templates from \citet{BC03} for Ellipticals, SB and Sc galaxies at redshifts $z<0.4$ are shown in cyan, green and yellow respectively. \emph{Bottom:} Colour-magnitude diagram $m_{\mathrm{F814W}}$ \emph{vs} ($m_{\mathrm{F606W}}-m_{\mathrm{F814W}}$) for galaxies within the ACS field without WFC3 imaging. The same colour code as above is applied. The bright source cut is shown as a dashed horizontal line. }
\label{fig:ccselec}
\end{center}
\end{figure}

\subsection{Weak-lensing constraints}

\label{sec:wl_constraints}
Here we present the details of the construction of the BUFFALO weak-lensing catalogue for Abell\,370.
The BUFFALO data offers a unique opportunity to model the mass distribution of Abell\,370 by combining both strong- and high resolution weak-lensing constraints. Thanks to \emph{HST}'s high resolution, we are able to construct a weakly-lensed background galaxy catalogue with a much higher density than what is feasible with ground-based data \citep[see for example][]{jauzac12,jauzac15a,jauzac16b,jauzac18b,medezinski13}.
To mitigate edge effects at the limits of the BUFFALO field-of-view, we combine the BUFFALO data with a weak-lensing catalogue obtained from Subaru/Suprime-Cam observations by \citet{umetsu2022}. We briefly describe this catalogue in this section. Both the BUFFALO and Subaru catalogues were aligned to the GAIA EDR3hj astrometry.

\subsubsection{The ACS catalogue}
We use the publicly available code \textsc{pyRRG} \footnote{\url{https://pypi.org/project/pyRRG/}} \citep[][]{harvey2019,harvey2021} on the ACS/F814W images to measure the shape of the background galaxies; these shapes carry the weak gravitational lensing distortion information. \textsc{pyRRG} is primarily based on a shape measurement algorithm first introduced by \cite{rhodes00}, RRG, where the shape of a galaxy is characterised using the second and fourth order normalised image multipole moments. Since RRG corrects for the Point Spread Function (PSF) on the raw image moments rather than the ellipticity, it avoids some uncertainties associated with other moments-based methods \citep[for more detail see][]{rhodes00}.

The \textsc{pyRRG} package initially extracts sources using a ``hot/cold'' method, whereby it separately uses \textsc{Source Extractor} \citep{bertin1996} with different size kernels to detect small and large objects in the observations, and then combines the catalogues. From the measured raw moments we select galaxies using a combination of the $\mu_{\rm max}$ and magnitude diagram, whereby we identify stars that will be used to measure the PSF, over-exposed stars that will require masking, galaxies that will be included in the final catalogue and objects that are too noisy to gain a reliable shape estimate.
From this catalogue, we first estimate the PSF for each {\it individual exposure} that comprises the final science image. We do this by comparing the moments of the stars within an image with a model of TinyTim \citep{tinytim} at different \emph{HST} focus positions. We find a best-fit PSF model and then iterate over all the exposures, stacking each PSF model until we have a final PSF for the science image. Using the final PSF model, we correct for the PSF, remove any double detections and sources within the light of massive cluster members or close to saturated stars, mask extra artefacts and produce a final estimate of the ellipticity. This method has been already tested and applied to previous cluster studies \cite[e.g][]{harvey15,jauzac15a,jauzac16b,jauzac18b,tam2020} and on galaxy-galaxy lensing \citep{harvey15b}. 

We apply additional cuts to clean the catalogue, by removing sources with potentially ill-measured shapes:
\begin{enumerate}
    \item sources located close to the edges of the different exposures by applying a cut on the number of exposures for each detected source;  
    \item objects with a size smaller than 0.11\arcsec, as they are smaller than the size of the PSF; 
    \item very faint sources, with a F814W magnitude cut at 28.5 in the HFF footprint and 27.0 elsewhere;
    \item very bright sources, with a cut at 23.5 in all fields; 
    \item we keep only objects with a well measured ellipticity, i.e. $0 < e < 1$, and with well measured errors on the shape measurement (which translates as errors > 0).  
\end{enumerate}

\subsubsection{Background galaxy selection}

The \textsc{pyRRG} output catalogue remains contaminated by foreground and cluster galaxies. Such contaminants will dilute the shear signal, it is thus necessary to identify them and remove from our lensing analysis. 
To calibrate the different selection cuts, we first use the photometric and spectroscopic catalogues presented in Sect.~\ref{sec:obs}. We define the cluster membership criterion as :
\begin{equation*}
    z_{\rm cluster} -dz < z < z_{\rm{cluster}}+dz,
\end{equation*}
where $dz=0.025$ and $0.04$ when considering spectroscopic and photometric redshifts, respectively \citep[spectroscopic cluster members are defined as having $z \in 0.375 \pm 0.025 $, see ][Fig. 3]{lagattuta2022}. 
Only $\sim 10\%$ of the sources in our catalogue have a photometric and/or spectroscopic redshift. Among those 10\%, $\sim 10\%$ are identified as foreground and $\sim 6\%$ as cluster members. 
To isolate background galaxies for the remaining 90\% of our catalogue that do not have a measured redshift value, we need to apply different colour and magnitude selections, depending on how many \emph{HST} filters they were observed with.

For the BUFFALO field of view observed with both ACS and WFC3, we apply a colour-colour selection following the method successfully applied by different weak-lensing teams \citep[e.g.][]{medezinski13,jauzac12}. Using the redshift information available, we identify the region populated by unlensed galaxies in the $m_{\rm F814W}$--$m_{\rm F606W}$--$m_{\rm F160W}$ colour-colour space and define its boundaries by:
\begin{align*}
(m_{\rm F606W}-m_{\rm F160W})&<1.3\times (m_{\rm F814W}-m_{\rm F160W})+0.8\ ;\\ 
(m_{\rm F606W}-m_{\rm F160W})&>2.3\times (m_{\rm F814W}-m_{\rm F160W})-0.6\ ; \\
(m_{\rm F606W}-m_{\rm F160W})&> 0.2\ . 
\end{align*}

We thus remove all objects within this region of the colour-colour diagram from our catalogue. The top panel of Fig~\ref{fig:ccselec} shows the $m_{\rm F814W}$--$m_{\rm F606W}$--$m_{\rm F160W}$ colour-colour diagram for Abell\,370, where the solid black lines show the colour boundaries defined here.
We validate this colour-colour selection by using colour predictions from spectral templates at the redshift of Abell\,370 and in the foreground. We combine empirical templates from \cite{CWW80} and \cite{kinney96} with theoretical ones from \cite{BC03}. Figure~\ref{fig:ccselec} shows the colour-colour tracks for several types of galaxies in the Hubble sequence at $z<0.4$, which agree well with our selection region highlighted by the black lines.

As the BUFFALO field of view is not imaged homogeneously by ACS and WFC3, not all galaxies in our catalogue have two colours. We thus cannot apply the above colour-colour selection to the entire catalogue. For the 1936 sources that only have $F814W$ and $F606W$ magnitudes, we apply a colour-magnitude selection, to remove cluster members.
This selection is described in more details in Sect.~\ref{sec:cluster_gals} as it is used to identify cluster galaxies included in our mass model, and we show the resulting selection in the bottom panel of Fig~\ref{fig:ccselec}. We note that the bright magnitude cut ($m_{\rm{F814W}} > 23.5$) also removes a large fraction of cluster member and foreground galaxies that appear below the red-sequence on Fig.~\ref{fig:ccselec}. We try more conservative cuts, by removing a larger region around the red-sequence and pushing the bright magnitude cut to higher magnitudes but it did not impact our results.

After removing galaxies located in the strong-lensing region of the cluster, the resulting sample of background galaxies contains 3581 objects, which gives a density of 42 sources/arcmin$^2$ in the HFF footprint, and 20 sources/arcmin$^2$ in the rest of the BUFFALO fields.

\subsubsection{The Subaru/Suprime-Cam catalogue}

We summarise briefly the construction of the Subaru weak lensing catalogue, but refer the reader to \citet{umetsu2022} for a detailed description. They measured the shape of the background weakly lensed sources on the $\sim 40 \times 40\,\rm{arcmin}^2$ Subaru/Suprime-Cam $R_{\rm{C}}$-band images
using their shape measurement pipeline based in part on the \textsc{imcat} package \citep[KSB,][]{kaiser1995}, incorporating key improvements developed by \citet{umetsu10,umetsu14}. To separate background from foreground/cluster galaxies, they applied a well-tested colour--colour selection using the Subaru/Suprime-Cam $BR_{\rm{C}}z'$ photometry. They selected two distinct populations of blue and red background galaxies identified in the $B - R_{\rm{C}}$ vs $R_{\rm{C}} - z'$ plane. The composite sample of blue+red background galaxies has a mean surface number density of $n_\mathrm{gal}\approx 21$~galaxies~arcmin$^{-2}$.

It is not within the scope of this paper to study the total matter distribution within the whole Subaru field. We only use this catalogue to mitigate possible edge effects that would come from abruptly cutting the model at the BUFFALO field-of-view. An analysis of the matter distribution within this larger field will soon be presented in Tam et al. (in prep.).

    \subsection{Cluster-member galaxy catalogue}
    \label{sec:cluster_gals}    

The parametric component of the model, described in \ref{sec:param_model}, contains haloes that account for the matter contained in cluster member galaxies and their associated dark matter.
We constructed a catalogue of cluster galaxies from the \textit{HST} mosaic as follows. 
\begin{enumerate}
    \item We ran \textsc{Source Extractor} on the full ACS/F814W and ACS/F606W bands mosaics in dual-image mode. The F814W band was used as the detection image, as it provides good sampling of the stellar population of elliptical galaxies at this redshift; the photometry was extracted from both images within the same aperture as measured in the F814W band. 
    \item Stars and non-astronomical artifacts were flagged and excluded based on their locus in the $\mu_{\rm{max}}$  vs magnitude  ($m_{\rm{auto}}$) diagram. 
    \item We cross-matched the coordinates of the resulting object catalogue with the MUSE spectroscopic catalogue from \citetalias{lagattuta19}. 
    \item We constructed a colour-magnitude diagram ($m_{\rm F606W}-m_{\rm F814W}$ vs $m_{\rm F814W}$) of the spectroscopic galaxies that are at the cluster redshift, i.e., $0.35<z<0.4$. We then fit the red sequence with a linear fit, using an iterative 3-sigma clipping process. This iterative process eliminates the blue cluster-member galaxies, and defines the spectroscopically-confirmed red sequence and measures its scatter. 
    We defined the red sequence as galaxies falling within a polygonal region in this colour-magnitude space, defined as 3-sigma above and below the linear fit, with the BCG F814W magnitude determining the bright limit, and the faint limit set to 25 mag.
    \item We selected the red-sequence galaxies from the entire BUFFALO field of view where F814W and F606W overlap, that are within the same red sequence box as defined from the spectroscopic sample. This forms the ``preliminary'' cluster member catalogue.  
    \item We then cleaned the preliminary catalogue by removing objects with MUSE spectroscopic redshifts that places them in the foreground or background. We checked the catalogue against objects identified as lensed arcs (however, no overlaps were found). Finally, a visual inspection was conducted by three of the authors to manually reject any remaining artifacts (mainly star spikes, edge effects) and other objects that were obviously not cluster galaxies (e.g., over-deblended emission regions in foreground galaxies).
\end{enumerate}

The final catalogue, containing 870 sources, tabulates for each cluster member galaxy its position (RA, Dec.), semi-minor and semi-major axis, position angle, and F814W magnitude values. The photometry was measured with \textsc{Source Extractor} as $m_{\rm{auto}}$. We add to this red-sequence-based cluster member catalogue, an additional 18 galaxies that are not located on the red-sequence, i.e. blue galaxies, but were identified as cluster members based on their spectroscopic redshifts. However, due to our cluster member selection technique, we do not include in our model blue galaxies outside of the cluster core. The fraction of blue to red galaxies is larger in the cluster outskirts than in the core \citep{lagattuta2022}, and they have in general a larger sub-halo mass at given galaxy luminosity, as they have been less subject to tidal stripping \citep[see Fig.~1 in][]{niemiec2022}. This could suggest a priori that we are missing a significant mass component with these galaxies, but is not necessarily true in practice. Even if they are more dark matter dominated compared to red galaxies, the blues are still on average less massive. In addition, the flexibility of our model outside of the cluster core allows to account for any ``missing'' mass component.

\subsection{Lightmap}
    \label{sec:lightmap}

In addition to the strong- and weak-lensing data described above, we make use of a cluster lightmap to analyse and interpret our results.
As lensing measures the total (dark+baryonic) matter distribution in the cluster, it is useful to estimate the distribution of the visible light in the cluster. We use for that the distribution of cluster member galaxies, and compute the corresponding lightmap in the following way: first we combine the BUFFALO cluster member catalogue (see Sect.~\ref{sec:cluster_gals}), with the cluster member catalogue extracted from the Subaru data \citep[extending to a larger radius, see][]{umetsu2022}, and the cluster members identified in the MUSE spectroscopic catalogue from \citet{lagattuta2022} (containing non-red-sequence galaxies). Using the \textsc{Source Extractor} segmentation map, we then extract the light corresponding to these galaxies from the Subaru-$z'$ image, and smooth it with a gaussian kernel with a width value of 35\arcsec. We tried different values for the size of the kernel, and found 35\arcsec to be a good trade-off, limiting the noise in the lightmap but still keeping some details in the light distribution.
We note that this lightmap only accounts for the contribution of the cluster member galaxies, and we therefore consider in our analysis that these are a tracer of the total mass distribution in the cluster. It neglects the contribution of the diffuse intra-cluster light, which represents between 1 and 10\% of the total cluster light for Abell\,370 \citep{montes2018}.

\section{Strong and weak-lensing mass modeling}
\label{sec:model}
In this section, we summarize the modeling methods used to study this cluster, which utilize the publicly available code \textsc{Lenstool} \citep{kneib96, jullo07, jullo09}.
\textsc{Lenstool} combines a ``parametric'' model that recovers the mass in the cluster core as being composed of a number of mass components whose positions and shapes are physically motivated, and a flexible grid model for the outskirts. The model is optimized with both strong and weak lensing constraints. After presenting the general method, we describe the different components in the mass model for Abell\,370.

Two optimization methods are tested here: (i) the Sequential Fit, where the cluster core is first modelled with the strong-lensing constraints, and then the outskirts are included in a second step with weak-lensing constraints; and (ii) the Joint Fit where both components are optimized jointly with strong- and weak-lensing constraints. We discuss the strengths and caveats of both methods at the end of the next subsection. In addition to these baseline models, results from which are presented in Sect.~\ref{sec:sequential_fit} and Sect.~\ref{sec:joint_fit}, respectively, we also try two alternative approaches to test our modeling choices: (i) a fully parametric model optimized with strong- and weak-lensing constraints, where substructures in the cluster outskirts are explicitly included as additional parametric potentials; and (ii) a fully non parametric model optimized with weak-lensing constraints only. The results of these models are presented in Sect.~\ref{sec:param_subs} and Sect.~\ref{sec:grid_only}, respectively, and the method implemented in these models is easily extrapolated from the baseline method presented in this section.

    \subsection{\textsc{Lenstool} strong- and weak-lensing modeling}
    \label{sec:lenstool_model}

In the core of galaxy clusters, the position of matter clumps can be \textit{a priori} estimated from the light distribution and the geometry of the strong lensing systems, which favours the use of a parametric approach to describe the total mass distribution in this region. This type of model consists typically of a small number of large scale potentials that reproduce the overall cluster mass distribution, and small scale potentials that account for the presence of observed cluster member galaxies.  We call $\bm{\Theta}$ the vector containing the free parameters describing these potentials, which are detailed in Sect~\ref{sec:param_model}. The total convergence field created by the parametric component of the model can then be written as:
\begin{equation}
    \kappa_{\mathrm{param}}(\theta) = \sum_{i} \kappa_{\mathrm{cluster}}(\theta, \bm{\Theta}_i) +  \sum_{j} \kappa_{\mathrm{galaxy}}(\theta, \bm{\Theta}_j),
\end{equation}
where $\kappa_{\mathrm{cluster}}$ is the convergence field corresponding to one large scale potential, and $\kappa_{\mathrm{galaxy}}$ to one cluster member galaxy. Throughout this paper, we call \emph{parametric model} such type of models, composed of a small number of physically motivated components.

In the clusters' outskirts a more flexible model is needed, as the positions of mass clumps are not known \textit{a priori} and the overall mass distribution can have a more irregular shape. As described in \citet{jullo09, jullo14}, we approximate the true convergence field $\kappa$ with a sum of Radial Basis Functions (RBFs) located on the nodes of an hexagonal multiscale grid,
\begin{equation}
    \kappa_{\rm{grid}}(\theta) = \frac{1}{\Sigma_{\rm{crit}}} \sum_i v_i^2 f(||\theta_i - \theta||, s_i, t_i),
\end{equation}
where $f(||\theta_i - \theta||, s_i, t_i)$, the RBF located at position $\theta_i$, with core and cut radii $s$ and $t$, is defined as:
\begin{equation}
	f(R, s, t) = \frac{1}{2G}\frac{t}{t-s}\left( \frac{1}{\sqrt{s^2 + R^2}} - \frac{1}{\sqrt{t^2 + R^2}} \right),
\end{equation}
and $v_i^2$ represents its weight. We denote $\bm{w}$ the vector containing the weights of all RBFs, which are the free parameters of the grid model. The values of the core and cut radii $s$ and $t$ are fixed for all the RBFs when the structure of the grid is set-up. The structure of the grid used in this analysis is described in Sect.~\ref{sec:grid_model}. We will refer to this type of modeling as \emph{grid}, \emph{free-form} or \emph{non-parametric models}. The total mass distribution is then modeled as the superposition of the parametric and grid components, $\kappa(\theta | \bm{\Theta, w}) = \kappa_{\rm{param}}(\theta | \bm{\Theta}) + \kappa_{\rm{grid}}(\theta | \bm{w})$.

Depending on the matter density in a given region of the cluster, this total mass distribution will deflect the light coming from background galaxies with varying strength. In the cluster core, lensing is strong, and multiple images of the same source can appear: the positions, $\theta_{\rm{I}}$, of the multiple images are thus used to constrain the mass distribution, as the position of an image can be expressed as
\begin{equation}
    \theta_{\rm I} = \theta_{\rm{S}} + \alpha(\theta_{\rm I}, \bm{\Theta}) + \sum_i v^2_i \rm{A}(||\theta_i - \theta_{\rm I}||, s_i, t_i),
\end{equation}
where $\theta_{\rm{S}}$ is the position of the corresponding source, $\alpha(\theta_{\rm I}, \Theta)$ is the deflection angle produced by the core parametric mass distribution at the observed image position, and $v^2_i \rm{A}(||\theta_i - \theta_{\rm I}||, s_i, t_i)$ is the deflection angle produced at the image location by the RBF located at position $\theta_i$ \citep[see for instance][for an analytical expression]{eliasdottir07}. At each step of the optimization process, for each system the observed positions of the multiple images are projected back into the source plane. As the model is imperfect, multiple images of the same system are not mapped back to the exact same source position. The barycentre of the different calculated source positions is therefore computed for each system, and the position of this barycentre is lensed back into the image plane.  This process, the ``image-plane optimization'' has the goal of minimizing the RMS distance between the calculated image positions and the observed ones. An existing alternative is the so-called ``source-plane optimization'' - in this case, the observed image positions are also projected back into the source plane and a barycentre is calculated, but the goal of the optimization is to reduce the distance between the individual source positions and that of the barycentre. Both methods has strengths and caveats, the latter are described in \citet{jullo07,jullo10}.

In the cluster outskirts, lensing is weak, and the observed images of background sources are only weakly distorted. The observed ellipticity of a source located at $\theta_{\rm I}$ can then be expressed as
\begin{equation}
\bm{e}_{\rm{obs}} = \bm{e}_{\rm{int}} + 2\bm{\gamma'}(\bm{\Theta}) +2\sum_i v^2_i \Gamma(||\theta_i - \theta_{\rm{I}}||, s_i, t_i),
\end{equation}
where $\bm{e}_{\rm{int}}$ is the intrinsic ellipticity of the source, $\bm{\gamma'}(\bm{\Theta})$, the shear produced by the parametric potentials, and $v^2_i \Gamma(||\theta_i - \theta_{\rm{I}}||, s_i, t_i)$ the shear produced by the RBF located at $\theta_i$. In this regime, the optimizations aim to find a mass distribution that would lens the background source population with a given shape noise into the observed image ellipticities.

The modelled mass distribution in the cluster is therefore constrained using these two sets of constraints: the position of the multiply imaged sources in the strong-lensing regions of the cluster, and the shape of the weakly lensed sources in the remaining regions. The likelihood function describing the model can then be written as
\begin{equation}
    \mathcal{L}(\bm{\Theta}, \bm{w} | \theta_{\rm{I}}, \bm{e}_{\rm{obs}}) = \mathcal{L}_{\rm{SL}}(\bm{\Theta}, \bm{w} | \theta_{\rm{I}}) \times \mathcal{L}_{\rm{WL}}(\bm{\Theta}, \bm{w} | \bm{e}_{\rm{obs}}),
\end{equation}
where the full expression for $\mathcal{L}_{\rm{SL}}$ and  $\mathcal{L}_{\rm{WL}}$ are given in \citet{niemiec2020}.

Ideally, the goal is to optimize the two components of the model jointly, using both strong- and weak-lensing constraints, as described above. This is the purpose of the recently developed \emph{hybrid}-\textsc{Lenstool} method presented in \citet{niemiec2020}, and what we refer to as \emph{Joint-Fit}. However, the \emph{hybrid}-\textsc{Lenstool} method still presents some computational limitations, the main one being that it is only able to perform the computation of the strong-lensing likelihood in the source plane, as opposed to the image plane. For complex clusters such as this one, the source plane optimization might not be sufficient to accurately recover the best fit for the mass reconstruction. We plan to test both methods on simulated clusters in future work. In the meantime, we show an attempt at a \emph{Joint-Fit} reconstruction in Sect.~\ref{sec:joint_fit}, but base our analysis on a more traditional \emph{Sequential-Fit}, where the parametric model in the core is first optimized with the strong-lensing constraints in the image plane, considering only $\mathcal{L}_{\rm{SL}}(\bm{\Theta} | \theta_{\rm{I}})$ (Sect.~\ref{sec:result_core}), and then fix this component to the best fit value, while the grid is optimized using the weak lensing constraints, $\mathcal{L}_{\rm{WL}}(\bm{w} | \bm{e}_{\rm{obs}}, \bm{\Theta}_{\rm{best}})$ (Sect.~\ref{sec:result_grid}).

    \subsection{Parametric model in the core}
    \label{sec:param_model}
    
In order to model the core of Abell\,370, we here use a similar mass decomposition to the \emph{copper} model in \citetalias{lagattuta19}:
\begin{itemize}
    \item 4 cluster-scale haloes to describe the overall mass distribution: one located around the position of the two BCGs (DM1 and DM3), one ``bridge'' halo that connects the two BCGs (DM2), and flattens the central mass profile, and a ``crown'' halo located in the North-East of the cluster (DM4); 
    \item the 2 BCGs of the cluster (BCG1 and BCG2), that are modelled separately from the rest of the cluster members, given that they are not expected to follow the same mass-to-light relation \citep[e.g.][]{richard10};
    \item 3 additional cluster member galaxies modelled outside the scaling relations (G1, G2, G3 and G4), that are located close to systems of multiple images, and therefore have a local impact on the geometry of the constraints;
    \item and a set of 449 cluster galaxies, identified within the entire BUFFALO footprint as described in Sect.~\ref{sec:cluster_gals}, and are modeled jointly following the scaling relations. Following \citetalias{lagattuta19}, we include in the lensing mass model  only cluster member galaxies with a magnitude $m_{\mathrm{F814W}} > 22.6$. 
\end{itemize} 
 
Each cluster-scale halo is modelled with a dual Pseudo Isothermal Elliptical mass distribution \citep[dPIE, see][]{eliasdottir07}, parametrized by its position, ellipticity, position angle, core radius and velocity dispersion values (the cut radius being fixed to 1000\,kpc). The BCGs and independent galaxies follow the same elliptical mass profiles but for them we fix the position, ellipticity and position angle to their observed light distribution, and only optimize the cut radius and velocity. The remaining cluster galaxies are modeled as follow: each is accounted for in the model by a dPIE matter component, but we reduce the number of free parameters by fixing their positions, ellipticities and angle positions to the \textsc{Source Extractor} measured values. In addition, we do not fit the remaining parameters for each galaxy individually, but only for a typical $L^{\star}$ galaxy ($m^{\star}_0 = 19.5$ in ACS/F814W).
The parameters of each galaxy are assumed to scale as:
\begin{equation}
\begin{cases}
	\sigma_0 = \sigma_0^{\star}\left(\frac{L}{L^{\star}}\right)^{1/4},\\
	r_{\rm{core}} = r^{\star}_{\rm{core}}\left(\frac{L}{L^{\star}}\right)^{1/2},\\
	r_{\rm{cut}} = r^{\star}_{\rm{cut}}\left(\frac{L}{L^{\star}}\right)^{1/2}.
\end{cases}
\end{equation}
We discuss the potential impact of tidal stripping on the satellite galaxies and their subhaloes in Appendix~\ref{sec:shmr}.
We refer the reader to \citetalias{lagattuta17} and \citetalias{lagattuta19} for more details on the construction of the parametric component of the mass model of this cluster and \citet{kneib&natarajan2011} for a more general discussion of substructure mass modeling.

To best reproduce the geometrical configuration, i.e., to obtain the lowest RMS separation between the model predicted and observed multiple image positions, \citetalias{lagattuta19} had to introduce an external shear component to their model.  
Such uniform shear fields, sometimes introduced in cluster strong lens modeling to improve the goodness of fit, produces a shear in the constraint position distribution. Usually, a shear field is held to account for tidal perturbations, generated by structures that are external to the modelled lens, either along the line-of-sight, or outside of the modelled field-of-view.  Alternatively, it can also compensate the lack of flexibility of a  parametric mass distribution, such as being restricted to elliptical potentials. 
This external shear component can therefore be considered as problematic for a comprehensive analysis of the total mass distribution, since it is not physically related to a mass component. \citetalias{lagattuta19} explored alternative models, by including some plausible background and foreground structures, but could not account for the shear component in this manner. For Abell\,2744, another massive merging cluster in the HFF/BUFFALO sample, it has been shown that massive substructures detected with weak lensing \citep{jauzac16b}, and located in the cluster outskirts, could impact the modeling of the cluster core, and explain a similar ``external shear'' component \citep{mahler18a}.
With this work, we aim at exploring whether that external shear component can be similarly removed,  by using weak-lensing constraints in order to precisely model the outskirts of Abell\,370. If massive substructures are present in the surrounding cluster environment, then they ought to create enough signal to be detectable with weak lensing, and we thus should be able to reproduce the \citetalias{lagattuta19} results while constructing an ``external shear-free'' model.

    \subsection{Free-form model in the outskirts}
    \label{sec:grid_model}

Outside of the strong lensing region of Abell\,370, we use a non-parametric grid to decompose the total matter distribution, ``on top'' of the parametric model described in Sect.~\ref{sec:param_model}. Indeed, it is important to note that the large scale mass components from the parametric model described above are not truncated at the limit of the strong-lensing region, but extend to the outskirts, which cause the parametric and grid models to spatially overlap. In this overlapping region, the total mass distribution is therefore a superposition of the mass contained in the parametric and grid models.
The grid is composed of RBF potentials (see Sect.\,\ref{sec:lenstool_model}) with fixed positions and sizes, parametrised by their core radius, $s$, and truncature radius, $t$. For all RBFs, the truncation radius is fixed at $t=3\times s$.   We use a multi-scale grid, whose resolution follows the density of the background weakly lensed sources. In the BUFFALO field of view, where deep \textit{HST} observations yield the highest source density, the grid is more resolved, while in the Subaru field we keep a low resolution to reduce the number of free parameters and the noise in the mass reconstruction. We also remove RBFs overlapping with the parametric component of the model in the cluster core. 

\begin{figure}
\begin{center}
\includegraphics[angle=0.0]{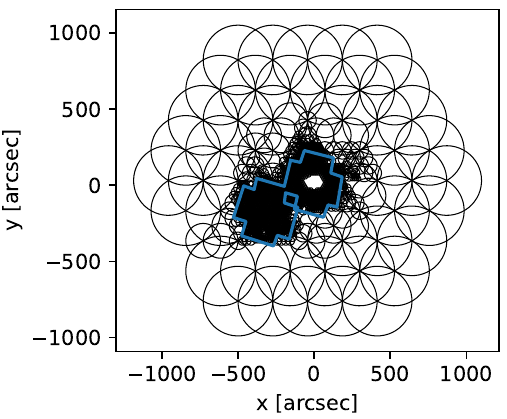}
\caption{Structure of the grid used for the free-form mass reconstruction. Each circle represents a RBF potential whose amplitude is optimized during the modeling process, and its size shows the core radius, $s$. The very dense region corresponds to the BUFFALO field of view, where $s=14''$, while in the Subaru field the resolution goes up to $s = 228''$. Potentials covering the cluster core region, modelled with the parametric part of the model, are being removed, which creates the white hole in the middle of the grid. The BUFFALO footprint is overplotted in blue, and the Subaru footprint extends to $\pm 1200\arcsec$ from the grid centre, covering the entire grid.}
\label{fig:grid}
\end{center}
\end{figure}

We test different grid resolutions, and obtain the optimum results with a grid composed of 1554 RBFs, with core radii $s$ ranging between 14\arcsec\ and 228\arcsec. The grid is created from a smoothed light distribution map tracing the background source distribution, using our publicly available set of scripts\footnote{\url{https://github.com/AnnaNiemiec/grid_lenstool}}. The structure of the resulting grid is shown in Fig.~\ref{fig:grid}.

\subsection{MCMC sampling}

The $(\bm{\Theta, w})$ parameter space is explored with 10 parallel Markov Chains, progressively converging from the prior to posterior PDFs following a variant of the Metropolis-Hastings algorithm \citep{metropolis1953, hastings1970} called selective annealing, using the publicly available sampler \textsc{BayeSys} \citep{skilling98}, which is implemented in the \textsc{Lenstool} algorithm. As described in \citet{jauzac12}, the amplitude of the RBFs is explored through Gibbs sampling by the \textsc{MassInf} extension of the \textsc{BayeSys} algorithm.

\textsc{BayeSys} samples the parameter space with 10 parallel chains, following two phases. First, a ``burn-in'' phase where the algorithm searches and converges towards the best-fit region of the parameter space, and then a ``sampling'' phase where it explores this region for a given number of MCMC steps. Following many previous analyses, here we fix this number to 100 steps, resulting in 1000 MCMC samples.

The parametric and grid models are based on different approaches towards the MCMC sampling: in parametric modeling, the goal is to find to set of parameters that best reproduce the observations. In this case, the ``output'' model is therefore the best-fit one, meaning the set of parameters, among the 10 chains $\times$ 100 steps realisations, that have the maximum likelihood. In contrast, the grid modeling approach is more statistical by nature, and instead of considering one realisation as the output model, we need to average the mass distribution with all the realisations of the sampled parameters. Throughout the paper, we will specify for the different considered models what we consider as the ``output'' model, whether the maximum-likelihood or the average over the 1000 samples. This combination of different modeling approaches may be one of the current limitation of our modeling, and is one of the aspects that we plan to study in details in a follow-up analysis performed on simulated clusters. In particular, in the case of the \emph{hybrid}-\textsc{Lenstool} grid+parametric combined model, we examine both the best-fit and the mean model, and find that even if they qualitatively agree, there are some differences in the value of the recovered parameters, and these differences need to be further examined in a future study.

\section{Results I: The total mass in A370 - Sequential-Fit model} 
\label{sec:sequential_fit}
\subsection{Modeling the core with strong lensing}
\label{sec:result_core}

The first step in our analysis is to create a ``baseline'' strong lensing model in the core of the cluster. We therefore optimize the parametric model described in Sect.~\ref{sec:param_model}, using the BUFFALO \emph{gold} strong lensing constraints, described in Sect.~\ref{sec:sl_constraints}. We start with broad and flat priors, and converge towards a best fit model using the \textsc{Lenstool} algorithm. We remind the reader that our baseline model is quite similar to the model presented in \citetalias{lagattuta19} in terms of the priors on the mass distribution decomposition. The main difference comes from the cluster member catalogue, as it was compiled within the BUFFALO collaboration independently from the previously existing  data set. However, the cluster member selection differences in the cluster core are marginal, the main difference being the extension of the galaxy catalogue towards larger cluster-centric distances, permitting better weak-lensing modeling. We note that our baseline model also contains the external shear component that was introduced in \citetalias{lagattuta19} to improve the goodness of fit of the model.

To quantify the goodness of fit of a strong-lensing model, the RMS distance between the observed multiple image positions and the positions predicted by the model is often used as a metric. The resulting RMS value for the best fit model is 0\farcs90, and the amplitude of the external shear component is $\Gamma = 0.107\,^{+0.002}_{-0.002}$, with an angle $\theta_{\Gamma} = -18.6\degree\,^{+0.2}_{-0.3}$ \citepalias[as compared to $\Gamma = 0.096\,^{+0.004}_{-0.003}$ and $\theta = -18.3\degree\,^{+1.1}_{-1.2}$, with RMS=0\farcs78 in the copper model from][]{lagattuta19}. This will be our baseline to evaluate the quality of alternative models, and probe how substructures in the cluster outskirts can account for (part of) the external shear. We present in Table~\ref{tab:model_summary} a summary of the external shear amplitude values and RMS for the different models considered throughout the paper.

\begin{figure}
\begin{center}
\includegraphics[width=\linewidth,angle=0.0]{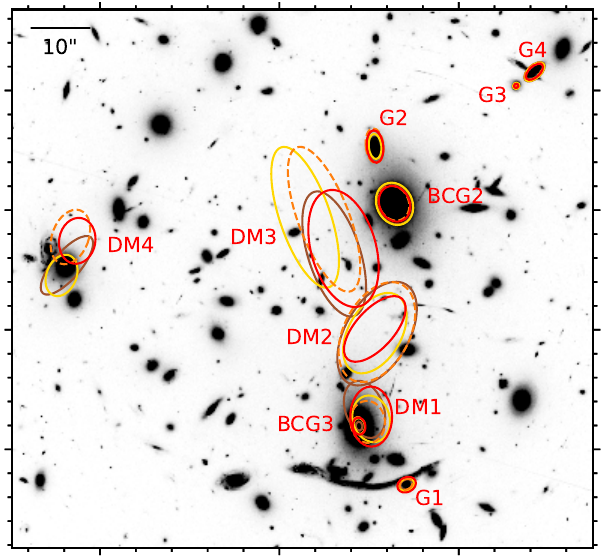}
\caption{Distributions and shapes of the mass components of the baseline parametric model in the cluster core, shown as red ellipses, with a nomenclature following \citetalias{lagattuta19} and reminded in Sect.~\ref{sec:param_model}. As a comparison, we show their \emph{copper} model with dashed orange ellipses. We also show as yellow ellipses the core model run with fixed substructures in the outskirts (Sect.~\ref{sec:fixed_subs}), and in brown with substructures optimized as parametric potentials (Sect.~\ref{sec:param_subs}). The size of the ellipses is a function of their velocity dispersion and cut radius, and therefore illustrates the relative mass of each halo.}
\label{fig:model_param}
\end{center}
\end{figure}

We show the distribution and shape of dark matter haloes (except the ones corresponding to cluster member galaxies modeled within the scaling relations) as red ellipses in Fig.~\ref{fig:model_param}: the four large scale haloes (DM1, DM2, DM3 and DM4), the two BCGs (BCG1 and BCG2), and the four independently modeled cluster members (G1, G2, G3 and G4). We show as a comparison the haloes comprising the best-fit \emph{copper} model from \citetalias{lagattuta19} as orange dashed ellipses. The shapes and relative contributions of the different constituent haloes vary between the two models, but the shape and amplitude of the overall reconstructed density profile is consistent with the models presented in \citetalias{lagattuta19}. We also note that the core mass distribution, and in particular the extra mass components DM3 and DM4 agree with recent non-parametric mass reconstruction presented in \citet{ghosh2021}.

\subsection{Modeling the outskirts with weak lensing}
\label{sec:result_grid}

\begin{figure*}
\begin{center}
\includegraphics[width=\textwidth,angle=0.0]{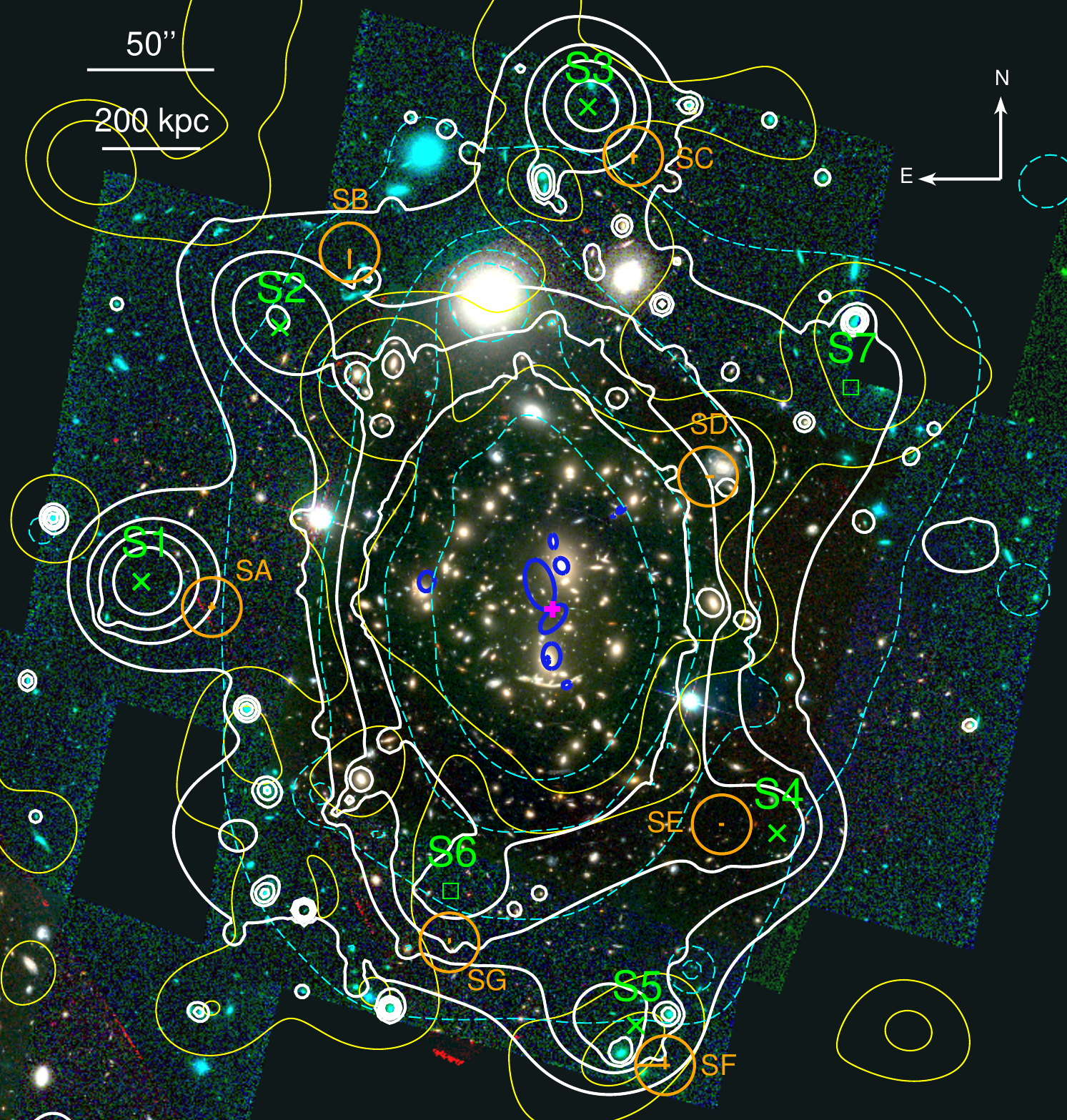}
\caption{BUFFALO colour-composite mosaic of the main field in the F160W, F814W and F606W filters. The yellow contours correspond to the lightmap derived from the BUFFALO+Subaru+Muse cluster member catalogue (see Sect.~\ref{sec:lightmap}); bold white contours show the lensing total mass levels; and the dashed cyan contours correspond to the \textit{XMM-Newton} X-ray surface brightness. The levels of the different contours where chosen arbitrarily to best illustrate the distribution of the different components. We show the position of the identified compact and extended substructures as green crosses (X) and squares respectively. For reference, we show the shape and position of the mass clumps composing the parametric model as blue ellipses in the cluster core. The orange circles, named SA,SB,SC,SD and SE, show the position of the parametric substructures optimized with both strong and weak-lensing constraints as described in Sect.~\ref{sec:param_subs}. The orange error bars within these circles represent the positional uncertainty for these substructures. he magenta cross indicates the reference centre RA$_{\rm{cen}} = 39.9706857$, Dec$_{\rm{cen}} = -1.5766997$. We note that the shape of the galaxy-scale potentials follow their light distribution, and is therefore elliptical for many galaxies, although they may appear more circular than they are due to the resolution of the figure.}
\label{fig:rgb_contours}
\end{center}
\end{figure*}

The second step in our analysis is to model the outer parts of the cluster (i.e outside of the core) using a free form matter decomposition as described in Sect.~\ref{sec:grid_model}, constrained with weak-lensing data as described in Sect.~\ref{sec:wl_constraints}. The matter distribution in the cluster core is fixed to the best fit model described in the previous Sect.~\ref{sec:result_core}. However, we note that we \emph{do not} include the external shear component in the model when we optimize the grid. This may add some inconsistencies into our model but it does not make sense physically to include a uniform external shear component that spans the entire cluster field. 
We examine some alternative models as well in subsequent sections of the paper that may also permit modelling the cluster more consistently at all scales. Our baseline \emph{Sequential-Fit} model therefore results in the combination of the parametric core model described in the previous section, and the free-form grid model described here. We refer to it as model$_A$. We note that we also run a grid model including in the core the \emph{copper} model from \citetalias{lagattuta19} instead of ours, to test the impact of this component on the mass distribution in the outskirts. We found no significant differences in the mass or inferred spatial distributions of the substructures.

\begin{figure}
\begin{center}
\includegraphics[angle=0.0]{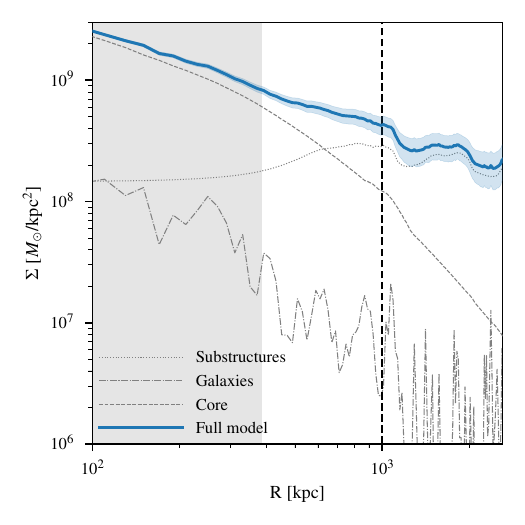}
\caption{Average surface mass density profile of the strong+weak lensing mass reconstruction for Abell\,370, with the centre taken at RA$_{\rm{cen}} = 39.9706857$, Dec$_{\rm{cen}} = -1.5766997$. The baseline model (Sect.~\ref{sec:result_core} and \ref{sec:result_grid}) is shown as the blue solid line. The blue shaded band represents the 1$\sigma$ statistical uncertainty on the total mass profile, computed as the 1$\sigma$ dispersion over the 1000 MCMC realisations. The different components of this model, i.e core, substructures and cluster members are shown in grey, with dashed, dotted, and dash-dotted lines respectively. We note that the core+galaxies density profile correspond to the parametric model described in Sect.~\ref{sec:result_core}, and the substructures to the grid described in Sect.~\ref{sec:result_grid}. }
\label{fig:density_profile}
\end{center}
\end{figure}

We present the total mass levels corresponding to the parametric+grid model$_A$ as cyan contours in Fig.~\ref{fig:rgb_contours}, overlaid on the BUFFALO main field colour-composite mosaic. The position and shape of the potentials composing the parametric best-fit model is also shown for information as blue ellipses located in the cluster core. To compare the weak-lensing mass distribution with the distribution of cluster members, we also plot the contours corresponding to the light distribution (see Sect.~\ref{sec:lightmap}) in yellow. We present in Fig.~\ref{fig:density_profile} the projected average surface mass density profile from this model, which is obtained by azimuthally averaging the lensing mass map, with the centre taken at RA$_{\rm{cen}} = 39.9706857$, Dec$_{\rm{cen}} = -1.5766997$. The solid blue line shows the mean profile, and the blue shaded region the standard deviation over the 1000 MCMC samples. We present the  density profiles corresponding to the different model components separately: the smooth core component, composed of the large scale parametric haloes (\emph{dashed line}), the grid-detected substructures (\emph{dotted line}), and the mass corresponding to cluster galaxies and their subhaloes (\emph{dashed-dotted line}). We highlight the strong-lensing region as a shaded grey area, which corresponds to the radial region containing all the strong-lensing constraints. We indicate the limit of the BUFFALO main field, the focus of this analysis, with a dashed vertical line.

From the weak lensing mass map, we identify in the BUFFALO main field seven candidate substructures S1, S2, S3, S4, S5, S6 and S7, all with a signal-to-noise (SN) higher than 3, where the SN is derived as follows:  SN maps are computed as the ratio between the mean and standard deviation maps from the 1000 MCMC realisations. We then compute the mean SN value within 175\,kpc from the centre of each substructure, to asses their significance over the 1000 realisations of the MCMC sampling. We do not apply a specific over-density finder algorithm, but identify them using iso-density contours on the lensing maps. This may cause our substructure identification to be somewhat arbitrary, and we plan to test in future works some dedicated algorithms to make this process more systematic. 

Among the seven candidates, we can qualitatively distinguish two classes of substructures: most of them are ``compact'', meaning that they correspond to only one RBF (+ eventual overlapping galaxy-scale PIEMDs). This is the case for S1, S2, S3, S4 and S5, and because of this property it is fairly easy to define the position of their centre. On the contrary, S6 and S7 are more extended, meaning that they are composed of a combination of multiple RBFs.  The positions of the centres of the possible substructures are shown as green crosses in Fig.~\ref{fig:rgb_contours}, for S1 to S5, and green squares for S6 and S7, to highlight the possibly higher uncertainty in their position: the mass levels show that S6 is  extending towards the East from the marked centre, with a possible secondary peak; the extended nature of S7 is not strongly visible on the mass levels, but it corresponds to a more diffuse mass overdensity extending towards the South-West of the identified centre of S7. We give distances of all the substructures, relative to the cluster centre located between the two BCGs (RA$_{\rm{cen}} = 39.9706857$, Dec$_{\rm{cen}} = -1.5766997$), in Table~\ref{tab:subs}: they are located relatively far from the cluster core, between $\sim 650$ and 1050\,kpc, which corresponds to $\sim 0.2 - 0.4\times R_{200}$. We note that these distances are only measured in projection in the sky plane, and the true 3D cluster-centric distances are probably larger. We measure the mass located within 175\,kpc from the identified centres of the candidate substructures, and present them in Table~\ref{tab:subs}. The total mass values are measured on the mass map, corresponding to the mean over the 1000 MCMC realisations, and the associated errors correspond to the standard deviation among them. We note that the quoted error only corresponds to the statistical error, and does not account for any systematic errors. If some substructures are only artefacts, for instance arising from noise in the weak-lensing catalogues, this would not be reflected in this error estimate.  

The masses of the substructures are each $\sim 6 \times 10^{13} \msun$ as measured from the grid+parametric mass map, which is in good agreement with for instance the mass distribution of substructures as measured in MACSJ0717 in \citet{jauzac18b}. However, it may be more significant to consider the overdensities these substructures represent, meaning the ``extra'' mass, as compared to the model containing only the smooth large scale components (i.e. the nine potentials from the parametric model). To quantify this, we measure the mass enclosed within 175\,kpc around the same positions, but on the mass maps corresponding to the core parametric model only.  We then give the substructure overdensities with respect to the parametric-only model as $\Delta M$ in Table~\ref{tab:subs}. The candidate substructures represent overdensities with masses between $\sim 3$ and $5 \times 10^{13} \msun$.\footnote{We measure the substructure over-densities with respect to the parametric model only as the grid RBFs contribute only marginally to the background cluster mass distribution: on a grid-only mass map, we measure the mass contained in random 175\,kpc apertures to be $\sim 10^8M_{\odot}$. } 

 \begin{table*}
    \centering
    \begin{tabular}{c|c c c c c c}
ID    & $M(<175 \rm{kpc})$   & $\Delta M$        & $M_{\star}(<175 \rm{kpc})$    & $M_{\star}/M$  & $R$    & $R$ \\
      & $[10^{12}\msun]$     & $[10^{12}\msun]$  & $[10^{12}\msun]$              & $[10^{-3}]$  & $[\rm{kpc}]$ & [\arcmin]  \\
\hline
S1  & $61\pm 10$  & $49 \pm 11$ & $0.08 \pm 0.01$ & $1.3 \pm 0.4$  & 863       & 2.70  \\ 
S2  & $62 \pm 9$  & $42 \pm 10$ & $0.01 \pm 0.01$ & $0.2 \pm 0.2$  & 821       & 2.57  \\ 
S3  & $60 \pm 10$ & $46 \pm 11$ & $0.17 \pm 0.03$ & $2.8 \pm 0.9$  & 1054      & 3.30  \\ 
S4  & $56 \pm 9$  & $34 \pm 10$ & $0.10 \pm 0.01$ & $1.8 \pm 0.8$  & 665       & 2.08  \\ 
S5  & $57 \pm 11$ & $35 \pm 12$ & $0.47 \pm 0.04$ & $8.2 \pm 2.3$  & 888       & 2.78  \\ 
S6& $71 \pm 9$  & $36 \pm 10$ & $0.28 \pm 0.02$ & $3.9 \pm 0.8$  & 629       & 1.97  \\ 
S7& $44 \pm 10$ & $26 \pm 11$ & $0.85 \pm 0.07$ & $19.3 \pm 6.0$ & 780       & 2.44  \\ 
    \end{tabular}
    \caption{List of substructure candidates. The total mass values are measured within 175\,kpc from the identified substructure centres, and correspond to the mean value over the 1000 realisations, while errors represent the standard deviations. We give the total mass values, $M$, as measured from the parametric+grid model, but also the masses overdensities, $\Delta M$, as compared to the parametric model only. $M_{\star}$ represent the total stellar masses within 175\,kpc estimated from the F814W magnitudes as described in Sect.~\ref{sect:sub_stars}, and $M_{\star}/M$ the stellar-to-total mass ratio as measured within 175\,kpc.   The distances are given relative to the cluster centre, located between the two BCGs (RA$_{\rm{cen}} = 39.9706857$, Dec$_{\rm{cen}} = -1.5766997$).}
    \label{tab:subs}
\end{table*} 
 
To verify whether the presence of these overdensities is simply a result of noise in the weak lensing catalogue, we perform a bootstrap analysis. First, we generate a hundred realizations of the weak lensing catalogue, each containing 80\% of the original catalogue, randomly selected for each realization. We then perform a mass reconstruction for each of these catalogues, and compute the mean mass map over the 1000 MCMC samples for each reconstruction, thus obtaining a hundred mass maps. We then compute the mean mass map over the hundred realizations, as well as the standard deviation map. We find that all of the 7 overdensities (S1 to S7) are detected in the mean mass map obtained from the bootstrap, and all have a signal-to-noise ratio higher than 3, where the SN is computed as the ratio between the mean and standard deviation maps. This gives us confidence that these overdensities are not a result of the presence of noise in the weak-lensing catalogue.  We note that this signal-to-noise is a different one from the one mentioned at the beginning of the section: the SN computed from the 1000 MCMC realisations accounts for the statistical noise at a given weak-lensing catalogue and model configuration, while this bootstrap checks the impact on substructure detection of possible correlated contaminants in the weak-lensing catalogue.  We will further verify the physical reality of these substructures by comparing the mass map with the distribution of cluster member galaxies and X-ray gas in the following sections, and then check the impact of our modeling choices on the presence and location of these substructures. We present the summary of these different tests in Sect.~\ref{sec:physical_reality}, and discuss there the physical credibility of each candidate substructure. To summarize, we find that out of the seven candidates, S1 and S4 are possibly modeling artifacts, and when considering only the most probable candidates, the cluster shows a mass distribution extended towards the North/North-West and the South-East.

\section{Results II: The total mass in A370 - Joint-Fit model} 
\label{sec:joint_fit}
As mentioned in Sect.~\ref{sec:lenstool_model}, the end goal is to model clusters self-consistently at all scales, using both strong- and weak-lensing constraints, following the \emph{hybrid}-\textsc{Lenstool} method presented in \citet{niemiec2020}. However, as introduced in Sect.~\ref{sec:lenstool_model}, this method is still under development, and for now only allows to optimize strong-lensing constraints in the source plane, which can lead to less precise reconstructions compared to image-plane optimizations. Due to this, \emph{hybrid}-\textsc{Lenstool} may not give yet, in its current avatar, the most precise reconstruction of the cluster core, but it is still an interesting method to verify if the inclusion of the substructures in the cluster outskirts can replace the external shear component as of now necessary in the Abell\,370 mass model. We therefore perform a Joint-Fit reconstruction of the cluster in the source plane, using \emph{hybrid}-\textsc{Lenstool}, and refer to this as model$_{D1}$ (we remind the reader that a summary of all the models presented in the paper is given in Table~\ref{tab:model_summary}).

\begin{figure}
\begin{center}
\includegraphics[width=\linewidth,angle=0.0]{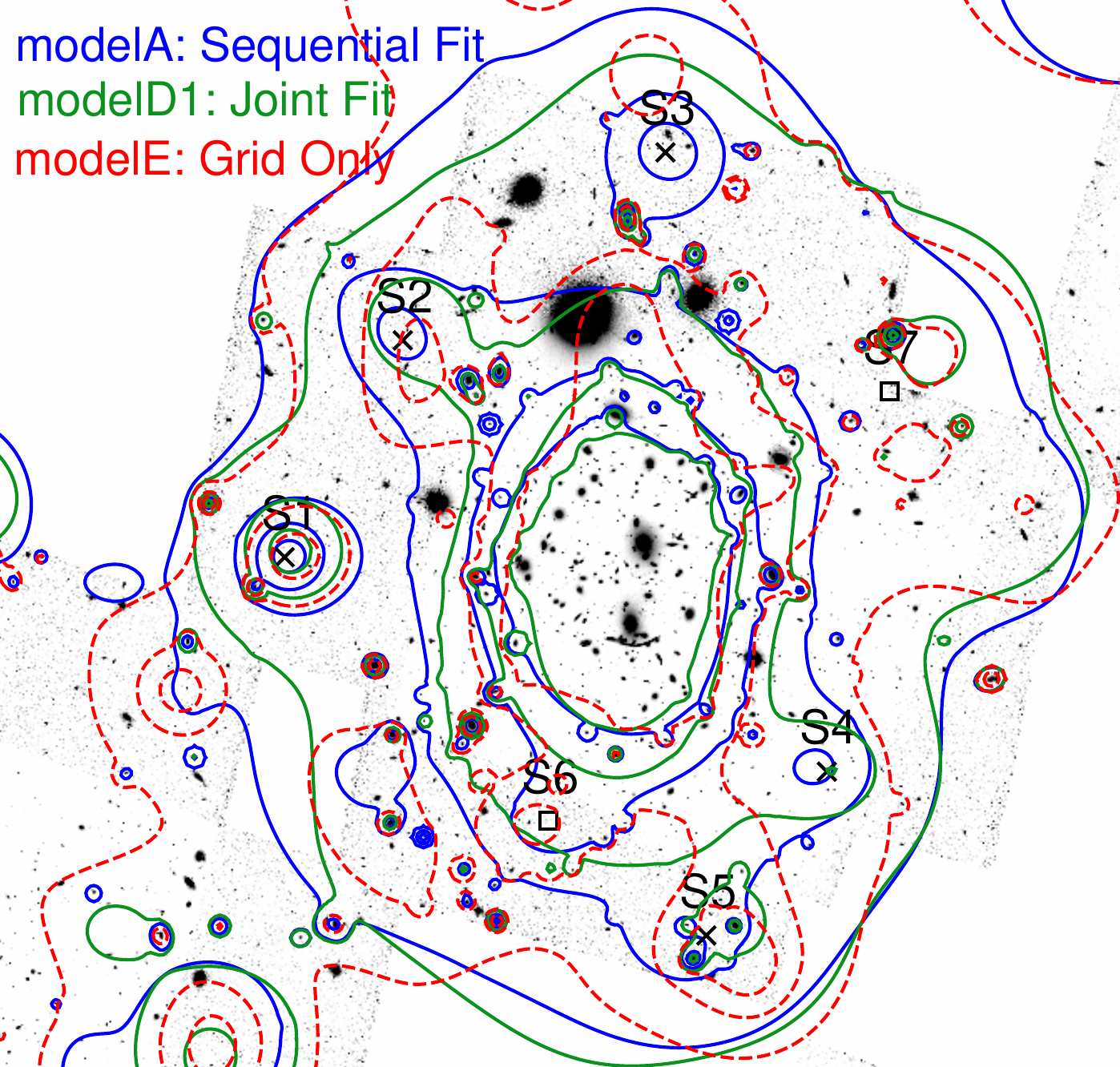}
\caption{Mass contours for different reconstructions: the Sequential Fit from Sect.\,\ref{sec:sequential_fit} in blue, the Joint-Fit from Sect.\,\ref{sec:joint_fit} in green, and the weak-lensing grid-only model from Sect.\,\ref{sec:grid_only} in red. The positions of the substructures detected in the Sequential Fit are also shown in black (as described in Sect.~\ref{sec:result_grid}).}
\label{fig:comb_map}
\end{center}
\end{figure}

The set-up of the model is the same as model$_A$ in Sect.~\ref{sec:sequential_fit}, with a parametric model in the core and a grid in the outskirts, but this time both components are optimised jointly using strong- and weak-lensing constraints. We note that we retain the external shear component in the parametric model, but leave broad priors on its amplitude. If the presence of substructures could replace this "artificial" component, the amplitude of the external shear should go to zero during the optimisation. The MCMC sampling phase is set again to 100 steps, with 10 parallel chains, and we generate 1000 mass maps. The output mass model is taken to be the mean over the 1000 maps, and we estimate the statistical errors on the reconstruction by computing the standard deviation map. We show the contours corresponding to the mean mass map in green in Fig.~\ref{fig:comb_map}.
The mass distribution is overall consistent with model$_A$, shown in blue, even if it differs slightly in shape or in amplitude. In fact, the same candidate substructures as in the Sequential-Fit (model$_A$) appear in the Joint-Fit (model$_{D1}$). We also present in green in Fig.~\ref{fig:all_density_profiles} the azimuthally-averaged surface mass density profile corresponding to this model, it is consistent with the Sequential-Fit model, within the statistical uncertainties.

The goal of implementing the Joint Fit is to examine whether the modeling of the core of the cluster can be improved by directly including the substructures in the outskirts. As mentioned above, this model was obtained with a source-plane optimization, and we therefore cannot directly compare it to the core-only strong-lensing model optimized in the image plane. We therefore re-run an optimization of the strong-lensing component of model$_A$, as described in Sect.~\ref{sec:result_core}, but this time performed in the source plane. This model yields an external shear amplitude $\Gamma_{\mathrm{SL, src}} = 0.109$, and a RMS$_{\rm{SL, src}}=1.56\arcsec$ (as compared to the image plane optimization which gives $\Gamma_{\mathrm{SL, img}} = 0.107$, and a RMS$_{\rm{SL, img}}=0.90\arcsec$).  For the Joint-Fit reconstruction performed in the source plane, the model with the total best likelihood has only a slightly lower external shear value, $\Gamma_{\mathrm{Joint, src}} = 0.090$, but a consistent goodness of fit, with RMS$_{\rm{Joint, src}}=1.57\arcsec$. All the RMS and $\Gamma$ values are summarised in Table~\ref{tab:model_summary} for comparison.
There are two main consequences to be drawn from this result: the first is the importance of the image-plane reconstruction as opposed to the source-plane one. The former allows to improve the precision and accuracy of the resulting total mass distribution in the cluster core. We are still developing this feature in \textit{hybrid-}Lenstool, and it should be available for future mass modeling efforts. Secondly, this measurement further underlines that Abell\,370 is a very complex cluster, and the current mass model studied here may be insufficient to fully characterize its mass distribution. In the Joint Fit, while the maximum-likelihood model has a poorer reconstruction of the core than the strong-lensing only model, this is not the best strong-lensing model among the 1000 model realizations. If we select the model with the best strong-lensing likelihood (regardless of the weak-lensing likelihood), it is actually better than what is obtained for the strong-lensing only model, with RMS=0.84\arcsec.

Finally, the external shear component included in the model makes this Joint-Fit very difficult to physically interpret. Even if the amplitude of the shear is reduced for the best-fit model, it is not completely removed.
Even if only considering the BUFFALO main field (and not the parallel, which is not discussed in this paper), this represents a region of $\sim 2 \times 2 \,\mathrm{Mpc}^2$. It is very difficult to physically account for such a uniform effect on a large scale (see Sect.~\ref{sec:toy_models} for a further discussion).
As a comparison, we re-run the same model but with the external shear component removed, but the resulting RMS is 1.84\arcsec (model$_{D2}$ in Table~\ref{tab:model_summary}). 
 
\section{Results III: The baryonic mass distribution in A370} 
\label{sec:baryons}
\subsection{Stellar content of substructures} 
\label{sect:sub_stars}
 
As described in Sect.~\ref{sec:sequential_fit} and \ref{sec:joint_fit}, we have detected 7 candidate substructures in the total mass distribution of the cluster, recovered using both the Sequential-Fit and Joint-Fit modelling methods. To gain a better sense of the physical reality of these candidates, we first quantify whether they correspond to overdensities in the cluster member galaxy distribution, and measure the total stellar mass contained within the same aperture. To estimate the stellar masses of galaxies, we follow the procedure outlined in \citet{jauzac15a}. We first estimate the typical $m_{\rm{F814w}} - m_{\rm{K}}$ colour for passively evolved galaxies at $z = 0.375$ using theoretical models from \citet{BC03}, assuming a range of exponentially decaying star formation histories within the range $\tau = 0.1-2\,\rm{Gyr}$. This gives a typical colour in the AB system $m_{\rm{F814w}} - m_{\rm{K}} = 1.3622$. We use this colour to compute $K$-band magnitudes for cluster galaxies from which we can then estimate the stellar masses, using the relation $log(M_{\star}/L_{\rm{K}}) = az + b$, where $z$ represents the redshift of the cluster, here $z = 0.375$. This relation was established by \citet{arnouts07} for red galaxies in the VVDS sample \citep{lefevre05}, adopting a Salpeter initial mass function (IMF), with parameters $a$ and $b$ given as:
\begin{align*}
    a = -0.18 \pm 0.03, \\
    b = -0.05 \pm 0.03.
\end{align*}
We thus estimate the stellar mass of all cluster galaxies detected in the BUFFALO field-of-view. As some substructures are located at the edge of this region, we perform the same exercise for cluster galaxies in the Subaru catalogue, and estimate their stellar masses from their $z'$ magnitudes.

\begin{figure}
\begin{center}
\includegraphics[angle=0.0]{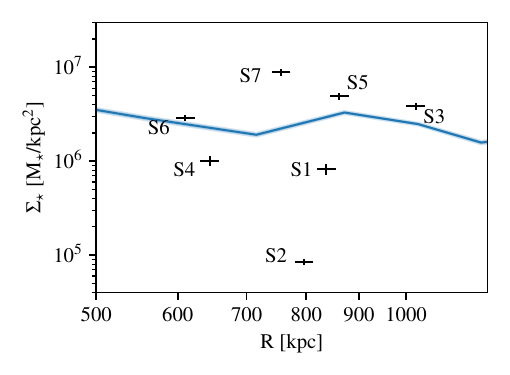}
\caption{Radially averaged stellar mass density profile for the cluster computed in circular bins (\emph{blue solid line}), and substructure stellar mass densities computed within $r=33\arcsec=175\,\mathrm{kpc}$. The length of the vertical lines represent the error propagated from the error on the F814W magnitude as given by \textsc{Source Extractor}. The same centre as in Fig.~\ref{fig:density_profile} is considered. }
\label{fig:rho_star}
\end{center}
\end{figure}

From the stellar mass estimates, we compute the total stellar mass included within 175\,kpc from the centre of the previously identified candidate substructures. We give these values in Table~\ref{tab:subs}. To give more credence to the physical reality of these candidate substructures, we would like them to match in positions with overdensities in the stellar mass distribution. This correlation can be examined qualitatively, using the lightmap and mass contours shown in Fig.~\ref{fig:rgb_contours}, but we use here the measured stellar masses to examine it quantitatively. For this, we measure the projected stellar mass density in each substructure, as $\Sigma_{\mathrm{sub}} = \frac{M_{\star}}{\pi r^2}$, where $M_{\star}$ is the total stellar mass within a substructure, and $r=33\arcsec$ (i.e., $175\,\rm{kpc}$). As a comparison point, we then compute the stellar mass density profile of the cluster, shown as a blue line in Fig.~\ref{fig:rho_star}.
We note that the stellar mass density profile is computed in circular bins, and do not account for the elongated shape of the cluster, which could make this comparison less meaningful.

Figure~\ref{fig:rho_star} shows that not all of the candidate substructures match with an overdensity in the stellar mass distribution: only S7, S5 and S3 have densities higher than the mean cluster density at their respective radius, which gives more credibility to their physical existence. 
Although S2  has the lowest stellar mass density of all, well bellow the cluster mean, there is a clear stellar overdensity located $\sim 0.76\arcmin$ South-West from the substructure's centre. This overdensity is too far to be accounted for with the total substructure stellar mass, but both can still be associated, as there may be some positional uncertainty associated with the detected candidate substructures. We briefly discuss these possible positional errors in Sect.~\ref{sec:grid_only}. Another possibility is that the detected overdensity could correspond to a trail of dark matter tidally stripped from the infalling group composed of the visible galaxy overdensity. The most striking discrepancy appears for the case of S1 and S4. These two substructures have no strong stellar counterparts, and their stellar densities are bellow the cluster mean. In the following sections, we examine X-ray observations of Abell\,370 to see if any gas counterparts for the different substructures can be identified, and we test alternative modeling methods in order to check whether these substructures can be artifacts of our modeling method.
 
\subsection{The X-ray gas}
\label{sec:xray_subs}

\begin{figure}
\begin{center}
\includegraphics[width=\linewidth]{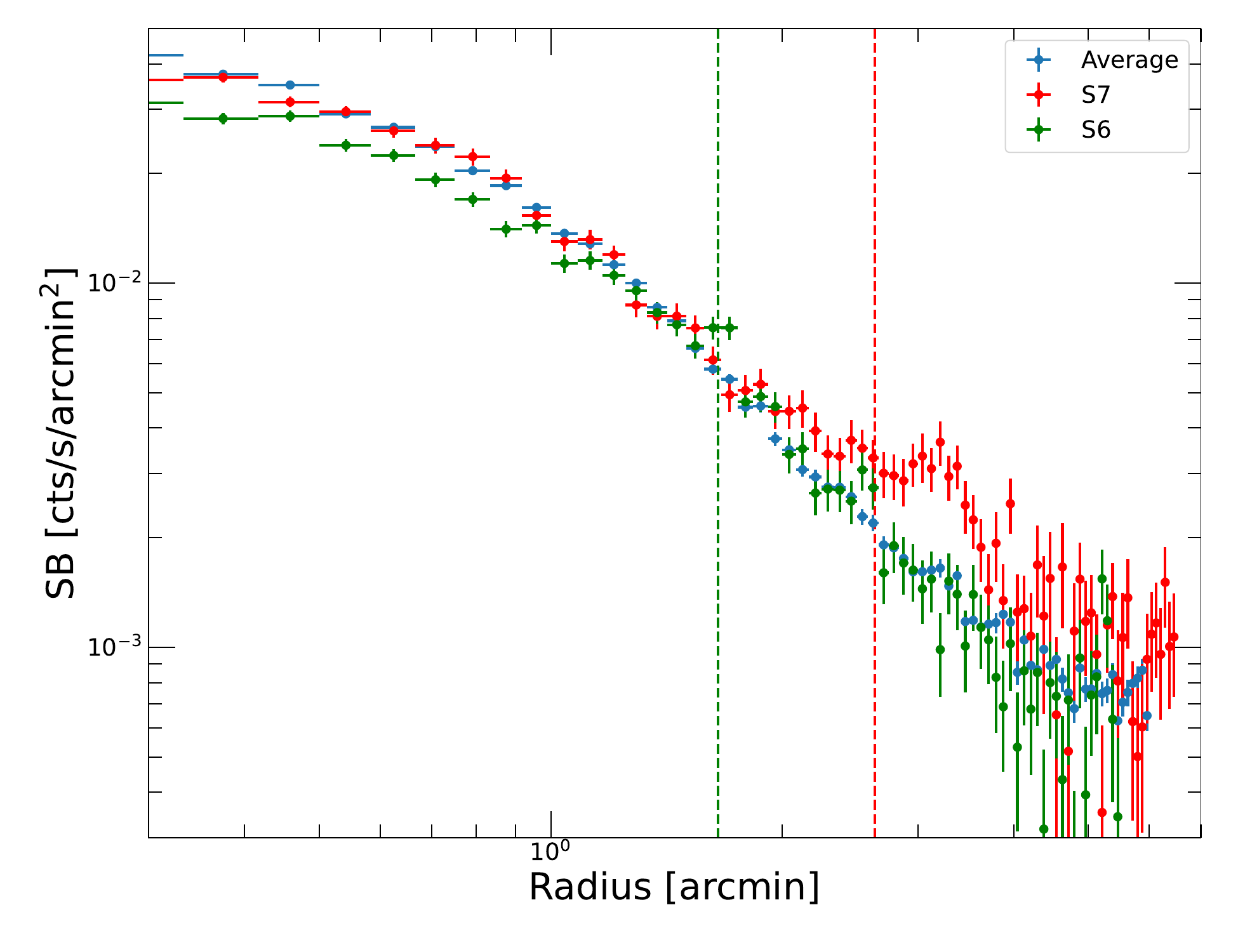}
\caption{\emph{XMM-Newton} surface brightness profiles of Abell\,370 in the [0.5-2] keV band. The average brightness profile of the cluster computed in circular bins is shown in blue, whereas the green and red profiles indicate the profiles extracted in boxes extending in the direction of the substructures S6 and S7, respectively. The dashed vertical lines indicate the distance of the centre of the S6 and S7 lensing structures to the cluster centre.} 
\label{fig:xray_profiles}
\end{center}
\end{figure}

To improve the interpretation of the lensing mass map and the candidate substructures, we analyse the X-ray observations presented in Sect.~\ref{sec:xray_data}. To quantify the presence of an X-ray counterpart at the location of the lensing substructures, we extract the X-ray surface brightness profiles using the public Python package \texttt{pyproffit} \citep{eckert20}. First, we measure the azimuthally-averaged profile of the entire cluster, by computing the mean luminosity in annular bins centered on the cluster core. This profile is shown in blue in Fig.~\ref{fig:xray_profiles}. The average profile is used as a baseline to search for surface brightness features in restricted regions of the cluster, corresponding to the directions in which each substructure is located. More precisely, for each substructure, we select a rectangular region, extending from the cluster center in its direction. We then compute the luminosity profile in rectangular bins within this box.
Figure~\ref{fig:xray_profiles} compares the azimuthally averaged cluster profile with the profiles calculated in the direction of S6 and S7, respectively. In the case of S7 (red profile in Fig.~\ref{fig:xray_profiles}) we observe a clear enhancement of surface brightness beyond 2 arcmin from the cluster centre, which is nicely consistent with S7's radial distance. The associated X-ray enhancement is obviously diffuse and extends over a broad radial range (2-4 arcmin) and the radial distance between the lensing position of S7 and the peak of X-ray luminosity is $\sim 0.6\arcmin$, as shown in Fig.~\ref{fig:xray_profiles}.  The brightness profile in the direction of S6 (shown in green) also exhibits a statistically significant brightness excess at the expected position, although the enhancement appears to be much more compact. We note that the profiles measured in the direction of S6 and S7 present lower values than the average profile in the central regions because they are measured in boxes of fixed width, while the average profile is measured in circular bins. For this reason, the average radius of pixels in the innermost radial bins is larger in the rectangular boxes than in the circular bins. For clarity, we do not show the profiles for any other candidates, as we do not find evidence for a surface brightness enhancement associated with any of the compact substructures. However, as already discussed in Sect. \ref{sec:xray_data} it is important to note that the Northern region of the cluster is affected by the presence of the foreground galaxy LEDA 175370, which is associated with a bright extended X-ray halo with a soft X-ray spectrum. While we mask a circle of 1\arcmin\ radius around the galaxy, any remaining extended X-ray emission from the galaxy could impact our measurements in the North direction, which may in principle affect our conclusions concerning S2 and S3.

\begin{figure*}
\begin{center}
\hbox{\includegraphics[width=0.5\textwidth,angle=0.0]{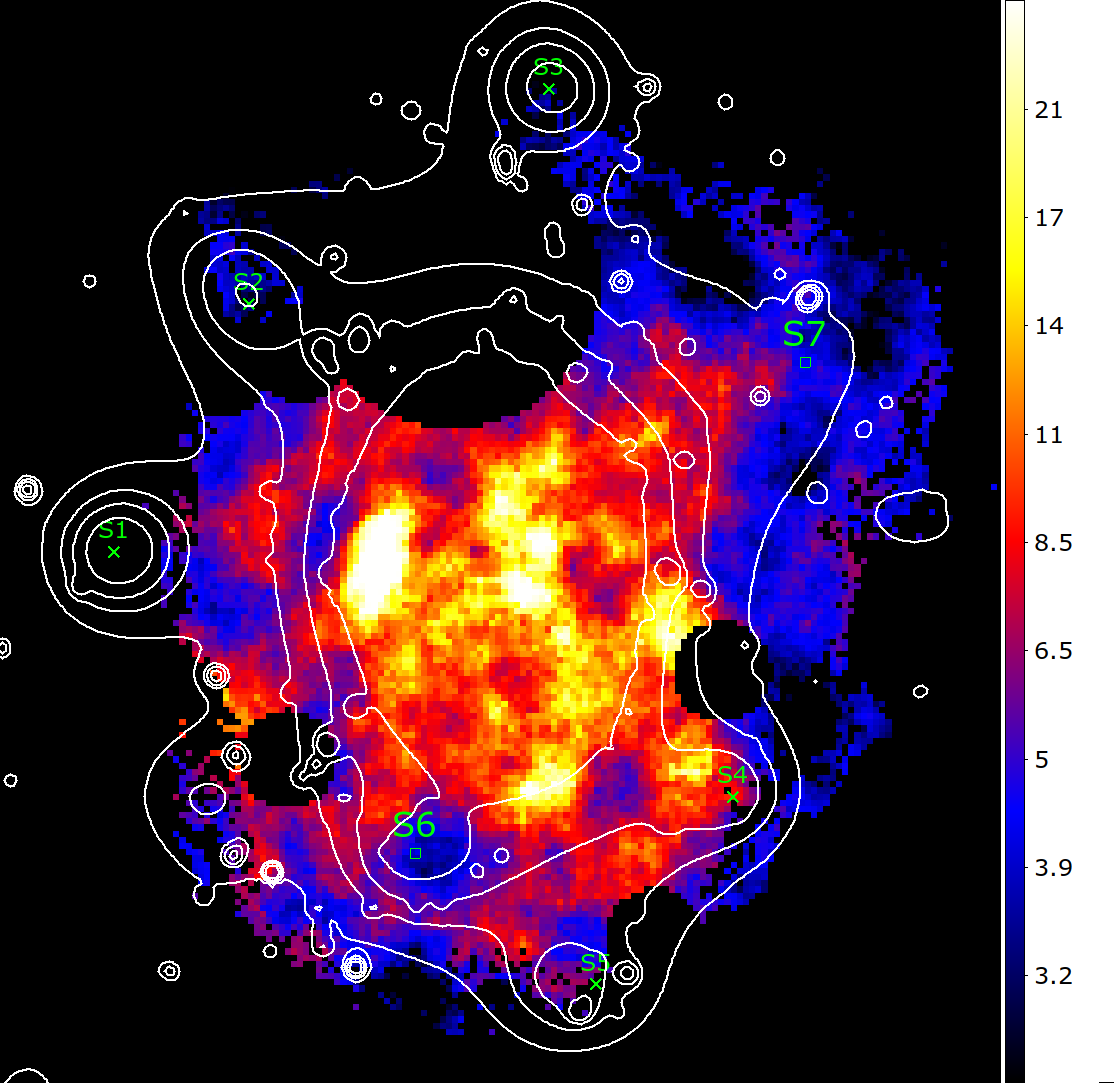}
\includegraphics[width=0.5\textwidth,angle=0.0]{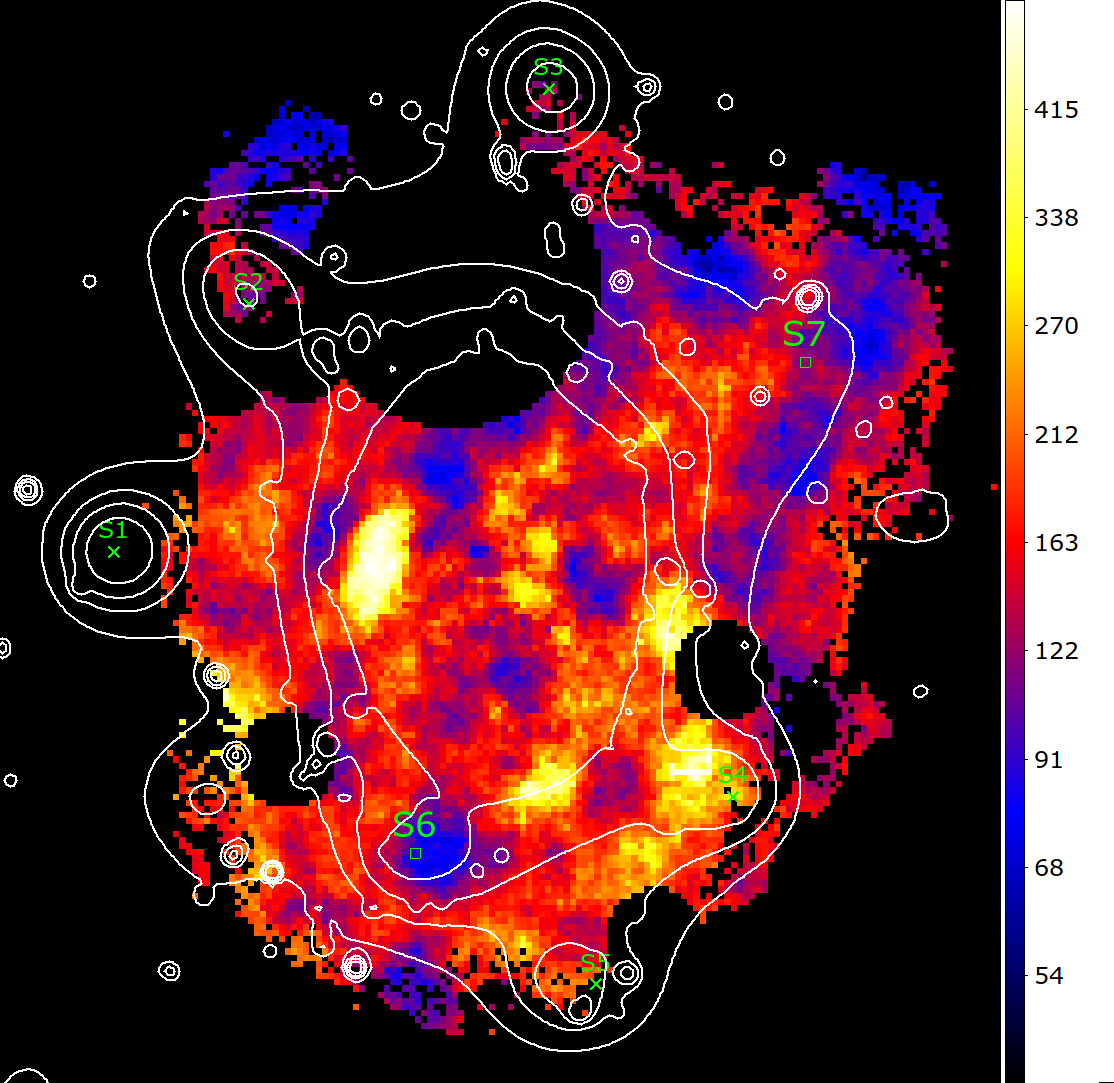}}
\caption{Maps of X-ray spectroscopic temperature (\emph{left panel}, in units of keV) and pseudo-entropy (\emph{right panel}, in units of keV cm$^{2}$). Details on the map construction are given in Sect. \ref{sec:xray_data}. In both panels, the white contours refer to the lensing contours (Fig. \ref{fig:rgb_contours}). The position of the lensing substructures is highlighted in green (see Table \ref{tab:subs}). 
}
\label{fig:xray_maps}
\end{center}
\end{figure*}

In addition, we show in Fig.~\ref{fig:xray_maps} the maps of X-ray spectroscopic temperature (\emph{left panel}, in units of keV) and pseudo-entropy (\emph{right panel}, in units of keV cm$2$). The position of S6 matches with a compact region with low temperature and low entropy, which would further suggest that S6 is a fairly recent infall, still containing some gas that has not virialized yet with the rest of the cluster. We discuss in more details the X-ray analysis of S6 in Appendix~\ref{sec:s6_xray}.

\section{Results IV: Impact of the substructures on the core model of A370} 
\label{sec:subs_impact}
\subsection{Fixed substructures}
\label{sec:fixed_subs}

In the previous sections, we have examined the candidate substructures detected in the total mass distribution of the cluster (Sect.~\ref{sec:sequential_fit} and ~\ref{sec:joint_fit}), and inspected their possible baryonic counterparts (Sect.~\ref{sec:baryons}). In this section, we verify whether this extra mass located in the cluster outskirts can have an impact on the lens model describing the cluster core.
This boils down to performing an additional step in the Sequential-Fit, meaning that we extract the potentials corresponding to the outskirt substructures, i.e. the RBFs from the grid model described in Sect.~\ref{sec:result_grid}, and keep them fixed while re-optimising the parametric model in the cluster core.  To do this, we first generate a \textsc{Lenstool} parameter file corresponding to the mean mass distribution over the 1000 MCMC parameter realisations obtained in our baseline model$_A$, and compute the average amplitudes for each grid potentials. We then include them as fixed potentials, and re-optimise the core strong-lensing model. We keep broad and flat priors for all the free parameters, but start from the best-fit values from Sect.~\ref{sec:result_core}. In particular, we take care of having broad enough priors on the amplitude of the external shear, so it can sample very low values if needed. We call this model$_{B1}$.

We note that we do not include all the grid potentials, in order to keep the model relatively simple. We only select grid potentials that can have a significant impact on the modelling of the core, i.e. all potentials with an amplitude $v > 50$, which results in 274 grid potentials. As described in the previous paragraph, these potentials are then kept in our model as fixed potentials (i.e., we do not re-optimise their amplitude). In order to reduce the computational time, we also fix the parameters of the scaling relation that govern the mass-to-light relation for cluster galaxies. 

We show the positions and shapes of the potentials in the core of the cluster corresponding to our best-model in Fig.~\ref{fig:model_param} as yellow ellipses. As could be expected, the presence of mass in the outskirts affects the model in the core of the cluster. The goodness of fit of the model is close to the core-only model: the RMS corresponding to this new model is RMS=0\farcs98. However, the external shear component is still necessary in this model, and cannot be replaced by these substructures: the best-fit model has only a slightly lower shear amplitude than in the core-only model, $\Gamma = 0.096$. This may be due to the fact that the candidate substructures are distributed along different directions, and therefore do not create a strong shear along a preferred axis.
To verify that the external shear is necessary in this configuration, we run again the model, but this time removing this component entirely. This model, called model$_{B2}$, is significantly worse, with RMS= 1\farcs42. We remind the reader that the different models presented throughout the paper are summarized in Appendix~\ref{sec:model_summary} along with the corresponding goodness-of-fit metrics.

\begin{figure*}
\begin{center}
\includegraphics[width=\textwidth,angle=0.0]{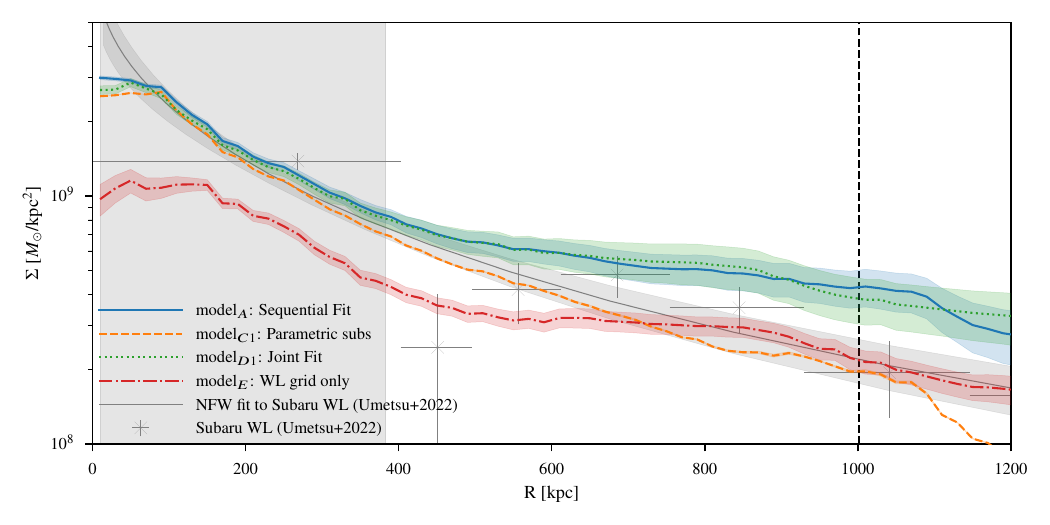}
\caption{Average surface mass density profiles for the different models considered: the baseline model in blue solid line (Sect.~\ref{sec:result_core} and Sect.~\ref{sec:result_grid}), the model with substructures included as parametric potentials in orange dashed (Sect.~\ref{sec:param_subs}), the source-plane joint fit in green dotted (Sect.~\ref{sec:joint_fit}), and the weak-lensing grid-only model in red dash-dotted (Sect.~\ref{sec:grid_only}). As a comparison, we present the density profile derived in \citet{umetsu2022} from the Subaru shear and magnification data as black marks with horizontal and vertical bars, and their NFW fit as a black solid line. The innermost bin is an integrated average inside $R=400$~kpc. The grey shaded area represents the strong-lensing region, and the dashed vertical line the limit of the main BUFFALO field.}
\label{fig:all_density_profiles}
\end{center}
\end{figure*}

\subsection{Parameteric strong + weak-lensing model}
\label{sec:param_subs}

We showed in the previous section that substructures in the outskirts, as detected with the baseline weak lensing analysis, cannot account for the external shear component that was introduced in the parametric model of the core. Here, we use another approach to test the impact of substructures on the cluster core. We model the  core with the same parametric model as in the baseline model$_A$, but we do not include the grid model in the outskirts this time. Instead, we add 7 parametric potentials that represent the substructures, initially distributed around the cluster, at the locations of the 7 substructures detected in the baseline model. We keep the same broad and flat priors for the core potentials as in Sect.~\ref{sec:result_core}. For the 7 new substructure potentials, we use circular dPIE mass distributions, and fix their core radii to $r_{\rm{core}} = 62$\,kpc, and cut radii to $r_{\rm{cut}} = 3\times r_{\rm{core}}$, which corresponds to the size of the most resolved potentials composing the free-form grid. For the remaining parameters, i.e. positions and velocity dispersions, we keep very broad priors, to give the model freedom to replace the external shear component with any mass and position of substructure it would take. Therefore, we let the velocity dispersion value vary between 20 and 700\,km/s for each substructure potential (which correspond to $7.2\times 10^{10}$ - $8.8 \times 10^{13}\msun$ in terms of total subhalo mass), and the width of the priors on the position is in the range 100-200\arcsec. We optimize this fully parametric model with both strong- and weak-lensing constraints, and refer to it as model$_{C1}$.

The resulting positions of the substructures are shown in Fig.~\ref{fig:rgb_contours} as orange circles, and are labelled as SA,SB,SC,SD, SE, SF and SG in order to facilitate the discussion. The final positions of these potentials are relatively close to the substructures identified in model$_A$ in Sect.~\ref{sec:result_grid}: SA, SB, SC, SE, SF and SG end up close to S1, S2, S3, S4, S5 and S6 respectively. The last one, SD, migrates closer to the cluster core. However, not all substructures end up with the same velocity dispersion value, and therefore not all have the same mass. We present in Table~\ref{tab:subs_slwl} the best-fit velocity dispersions for the 7 substructures, as well as the total mass measured on the resulting mass maps within 175\,kpc from their centre. As in Sect.~\ref{sec:result_grid}, we also compute for each of them $\Delta M$, the excess mass enclosed in the same region with respect to the core only model. 

This reveals that only 3 substructures correspond to a large amount of additional mass: SC, SE and SG represent additional mass components of $2.9 \pm 0.2$, $2.2 \pm 02$ and $3.3 \pm 0.2 \times 10^{13}\msun$ within 175\,kpc respectively, which could represent the mass of small galaxy groups. SC is located close to an overdensity in the cluster member light distribution, and while it is not exactly co-spatial, this small shift could be attributed to systematic uncertainties in the measurements of substructure positions. This is reinforced by the fact that SC is located at $\sim 30\arcsec$ from the substructure S3, as measured on the grid mass reconstruction. We plan to address this issue of systematic errors on the reconstructed positions of substructures in a future work on simulated clusters. In contrast, SE is not located close to any overdensity in the cluster baryonic mass distribution, neither traced by optical or X-ray emissions. We tend to classify it as a modeling artefact, but it is still puzzling that it is present both in the grid, and in the parametric-only mass reconstruction with large enough mass attributed to it. Finally, SG is located fairly close to S6 ($9\arcsec$ distance), and to the extended cluster member over-density located to the East. Similarly as for SC, the question of systematic uncertainty in the substructure's reconstructed position arises. Next, in decreasing order of mass, is SD, with an excess mass of $1.1 \times 10^{13}\msun$. This halo presents the largest positional shift as compared to its initial position, and it ends-up coinciding with the position of a foreground galaxy at $z=0.25$. This could suggest that this line-of-sight structure may contribute to the total lensing efficiency. Overall, the substructures modeled as parametric potentials are all fairly grouped along the North-South axis of the cluster, confirming that it is the main elongation axis of Abell\,370. We also show the statistical uncertainty on the recovered position of the substructures, $1\sigma$, as orange crosses in Fig.~\ref{fig:rgb_contours}.  We now examine if and how the model in the core is modified in this configuration.

\begin{table}
    \centering
    \begin{tabular}{c|c c c}
        & $\sigma_{\rm{LT}}$& $M (< 175\rm{kpc})$   & $\Delta M (< 175\rm{kpc})$    \\  
        & [km/s]            & $[10^{12}\msun]$      & $[10^{12}\msun]$  \\
    \hline
     SA  & $287 \pm 10$     & $27 \pm 1$            & $9 \pm 2$         \\  
     SB  & $234 \pm 9$      & $27 \pm 1$            & $5 \pm 2$        \\   
     SC  & $547 \pm 4$      & $46 \pm 1$            & $29 \pm 2$         \\  
     SD  & $287 \pm 8$      & $59 \pm 1$            & $11 \pm 2$         \\ 
     SE  & $515 \pm 9$      & $54 \pm 1$            & $22 \pm 2$        \\  
     SF  & $224 \pm 5$      & $26 \pm 1$            & $8 \pm 2$        \\  
     SG  & $562 \pm 15$     & $61 \pm 1$            & $33 \pm 2$        \\  
    \end{tabular}
    \caption{List of substructure velocity dispersions and masses, resulting from the strong+weak-lensing parametric model. The masses are measured on the mass maps within 175\,kpc from the identified substructure centres, and correspond to the mean value over the 1000 realisations, while the error represents the standard deviation. We also give the mass overdensities, $\Delta M$, with respect to the core only parametric model. }
    \label{tab:subs_slwl}
\end{table} 

The positions and shapes of the potentials comprising the core matter distribution are shown as brown ellipses in Fig.~\ref{fig:model_param}. This model appears quite similar to the baseline core model, but, in this configuration, substructures actually account for a significant fraction of the external shear amplitude: in the best-fit model its amplitude is reduced to $\Gamma = 0.04$, which is about a third of its initial value. However, this component is still present, and this makes this model still not completely satisfying from a physically motivated point of view: there is a uniform shear component covering the whole modeled field, $\sim 4' \times 4'$, which is difficult to reproduce by adding mass components (see Sect.~\ref{sec:toy_models}). We note that this model is slightly worse at reproducing the observational constraints, as its RMS is 1.17\arcsec. We include the RMS and $\Gamma$ values in Table~\ref{tab:model_summary} along with the other models. Finally, we compute the azimuthally averaged density profile for this model, shown in Fig.~\ref{fig:all_density_profiles} as orange dashed line. It shows that outside of the core, this model yields a lower average surface mass density than the baseline model$_A$: this is due to the fact that the outskirts mass is concentrated in a small number of localized substructures, instead of being allowed to follow a much smoother distribution, as with the grid-type  models.

As a comparison, we also perform this modelling exercise without including the external shear component, to see if this will force more mass into the substructures, and perhaps still manage to give a decent model, which we call model$_{C2}$. The positions and masses of the substructures are different in this configuration. 
Out of the 7 substructures, 5 have now masses $M(< 175 \rm{kpc}) > 10^{13} \msun$, and the remaining two contribute only marginally to the total mass distribution ($M(< 175 \rm{kpc}) \sim 5\times 10^{12} \msun$). Their positions remain within $15\arcsec$ of the positions obtained in the former model, with one notable exception: substructure SA migrates to the South-West, and its best-fit position coincides with the group of 3 relatively large cluster member galaxies located between S1 and S6 in Fig.~\ref{fig:rgb_contours}. In this configuration, the substructures SA and SG are located within $25\arcsec$ of each other, and are the two most massive. Together they contain $> 10^{14} \msun$, which corresponds to the mass of a small galaxy cluster. However, there is no clear luminous counterpart for such a massive structure at this location. We believe that the very high mass of these substructures is rather a model artifact, driven by the lack of external shear.
In this model configuration, the distribution of the substructures along the North-South axis is even more pronounced, which would also tend to produce a stronger total shear on the core (see Sect.~\ref{sec:toy_models}). As discussed previously, this model may not be entirely physically motivated, due to the presence of extremely massive substructures lacking a clear optical counterpart, but it does manage to replace the external shear component, and obtain a good reconstruction of the position of the multiple images.  The RMS value is here equal to 1.19\arcsec, which is equivalent to the same model with the external shear component. We remind the readers that the RMS only reflects the accuracy of the model in the strong-lensing region, but it is the appropriate metric here, as in this section we are examining the impact of the presence of substructures on the model in the cluster core.

\section{Discussion}
\label{sec:discussion}

\subsection{External shear toy models}
\label{sec:toy_models}

One purpose of this paper is to study the physical origin of the external shear component that is required to improve the goodness of fit of the lens model in the core of Abell\,370. The amplitude of the required shear is quite high in our baseline model ($\gamma \sim 0.11$), and we show in Sect.~\ref{sec:subs_impact} that the substructures detected in the cluster outskirts can only account for a small fraction of it. In this section, we aim at quantifying if the presence of (a) physically realistic substructure(s)  could produce an external shear of this amplitude, by using a simple toy model. For this, we simulate the presence of one or two substructures in the outskirts of a cluster, and generate a 2D map of the shear it produces in the cluster field, and thus the value of "external" shear it generates at the position of the cluster core. We vary the position of the substructure(s) and measure the resulting variations of the shear amplitude in the cluster core. 

\begin{figure}
\begin{center}
\includegraphics[angle=0.0]{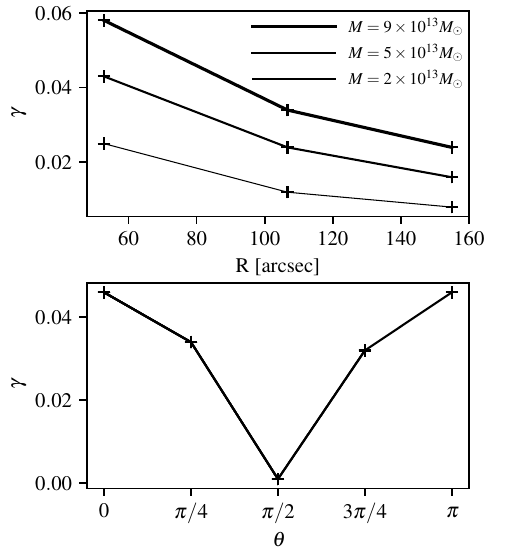}
\caption{\emph{Top panel:} In the case of one substructure, amplitude of the shear created in the cluster core as a function of the substructure distance to the core. Lines with increasing thickness represent a substructure with $M_{200} = 2,\,5\,\mathrm{and}\,9\times 10^{13}M_{\odot}$ respectively. \emph{Bottom panel:} In the case of two substructures, amplitude of the shear as a function of the angle between the two substructures. The two substructures are located at 107\arcsec = 570\,kpc from the cluster centre, and have $M_{200} = 5\times 10^{13}M_{\odot}$.}
\label{fig:toy_model_gamma}
\end{center}
\end{figure}

\begin{figure*}
\begin{center}
\includegraphics[angle=0.0]{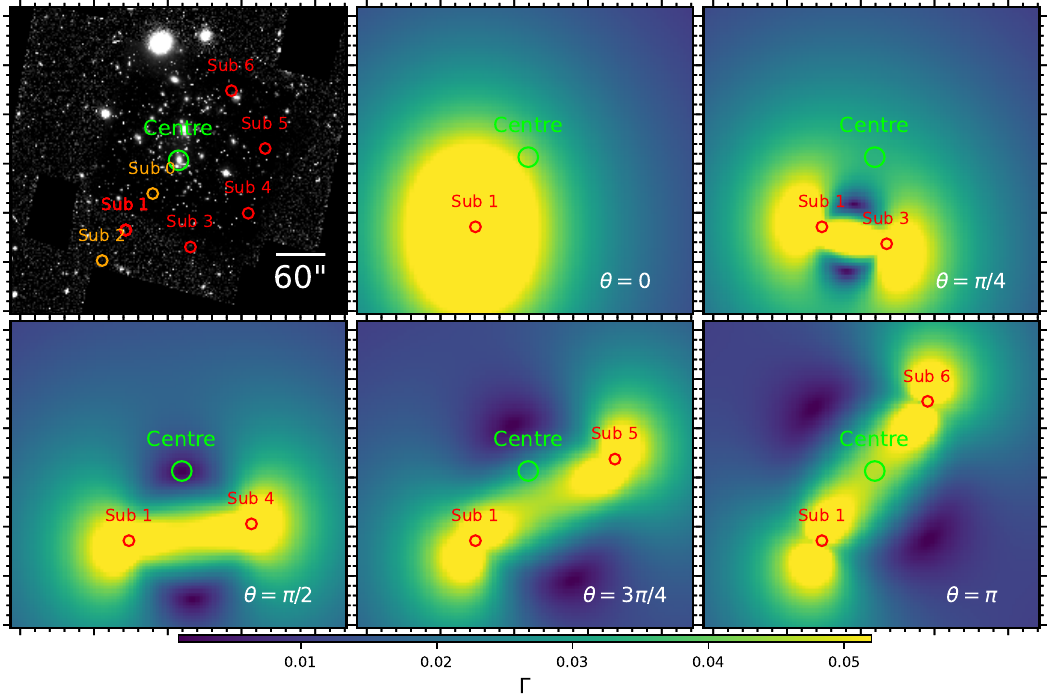}
\caption{\emph{Top left:} BUFFALO main field for Abell\,370, with the positions of the different substructures considered in the toy models shown as red and orange circles. The cluster centre position is shown as a green circle. \emph{Remaining panels:} 2D maps of the shear generated by two substructures with $M_{200} = 5 \times 10^{13}M_{\odot}$ and located at 107\arcsec = 570\,kpc, with different angular configurations. The color map represent the shear generated by the substructures at each position in the field, yellow (blue) representing high (low) values of the shear. The positions of the substructures were chosen to sample the parameters that impact the amplitude of the shear in the cluster core (i.e distance to the cluster centre, relative angle between the substructures). We do not examine here the orientation of the shear field, which would be impacted by a rotation of the substructure configuration for instance.}
\label{fig:toy_model_gamma_2d}
\end{center}
\end{figure*}

We start by including only one substructure, modeled with a NFW profile \citep[][]{NFW96}, and with a mass $M_{200} = 5\times 10^{13}M_{\odot}$, which is a typical mass for group-like substructures that can be detected in cluster outskirts \citep[e.g.][]{harvey14, jauzac15a}.
We consider different distances between the position of this single substructure and the core of the cluster, from $\sim 50$\arcsec\ to $\sim 150\arcsec$, which corresponds to the positions labeled Sub 0, Sub 1 and Sub 2 on the top left panel of Fig.~\ref{fig:toy_model_gamma_2d}. For each of these substructure positions, we measure the amplitude of the shear generated at the cluster core.
The resulting values are shown as a function of the distance to the cluster centre in the top panel of Fig.~\ref{fig:toy_model_gamma} as the medium thickness line. As expected, the closer the substructure, the stronger the external shear. We also test this setup with different substructure masses, $M_{200} = 2\times 10^{13}M_{\odot}$ and $M_{200} = 9\times 10^{13}M_{\odot}$, shown with thin and thick lines respectively. 
The top panel of Fig.~\ref{fig:toy_model_gamma} shows that one substructure is not enough to explain the external shear needed in  Abell\,370, as even the most massive considered substructure, located the closest to the cluster core only accounts for around half of the expected shear. We note that having such a massive substructure ($M_{200} = 9\times 10^{13}M_{\odot}$) located so close to the cluster core ($\sim 53$\arcmin\ or 277\,kpc) is not realistic, as it should be detected with weak lensing and/or galaxy over-density.

As a second step in the toy model, we now include two identical substructures with NFW profiles, and masses $M_{200} = 5\times 10^{13}M_{\odot}$. The two substructures are located at 107\arcsec = 570\,kpc, and we vary the angle between the two substructures (considering the cluster core as the center): $\theta = 0$ represents when the two substructures are superposed, resulting in one substructure twice as massive, and $\theta = \pi$ represents where the two substructures are opposed with respect to the cluster centre (see Fig.~\ref{fig:toy_model_gamma_2d} for the relative positions of the substructures and the cluster centre). The bottom panel of Fig.~\ref{fig:toy_model_gamma} shows the resulting amplitude of the "external" shear measured in the cluster core . The shear is maximal when the two substructures are superposed (position Sub 1 on Fig.~\ref{fig:toy_model_gamma_2d}) or aligned with the cluster center (Sub 1 and Sub 6 on Fig.~\ref{fig:toy_model_gamma_2d}), as both substructures contribute to the shear in the same direction. In contrast, it cancels out when they are perpendicular (Sub 1 and Sub 4 on Fig.~\ref{fig:toy_model_gamma_2d}), as in this case each substructure distorts the background galaxies in orthogonal directions. By comparing the top and bottom panels of Fig.~\ref{fig:toy_model_gamma}, we also note that having two substructures (superposed or aligned) located at $\sim 107$\arcsec\ from the cluster core (bottom panel, $\theta = 0$ or $\pi$) is equivalent to having one substructure with the same mass but located half closer (top panel, medium thickness line, $R \sim 50\arcsec$ ).

In conclusion, it seems difficult to explain the external shear necessary in the core of Abell\,370 with the presence of substructures, as it would require two very massive substructures ($M_{200} \sim 9\times 10^{13}M_{\odot}$) aligned with the cluster core, and located quite close ($\sim 50$\arcsec\ or 270\,kpc). Given the orientation of the external shear component, they should be roughly located along the axis defined by the positions Sub 0, 1, 2 and 6 in the top left panel of Fig.~\ref{fig:toy_model_gamma}. Our mass reconstruction (see Sect.~\ref{sec:result_grid}) does not show such massive substructures.  

Finally, we examine the 2D maps of the shear amplitude generated by the two-substructures toy models. Figure~\ref{fig:toy_model_gamma_2d} shows the shear maps corresponding to the five configurations which measurements are given in the bottom panel of Fig.~\ref{fig:toy_model_gamma}. The positions of the substructures are shown as red circles in each panel, and the position of the cluster centre as a green circle. Depending on the substructure positions, the distribution of the shear takes different patterns, but is never uniform in the cluster field. It is therefore difficult to physically account for an "external" shear component by including substructures, as in the parametric model, this is a uniform component over the whole field. The external shear components often used in parametric modeling of galaxy clusters are of course approximations, as are all model decompositions, but they should be treated with care when it comes to their physical interpretations: they can be an approximation for the impact of some (sub)structures, but can also be the result of other approximations in the modeling, such as the limited choice in terms of potential shapes.

\subsection{Weak-lensing grid-only model}
\label{sec:grid_only}

We have shown that the complexity of Abell\,370 leads to several difficulties when it comes to producing a strong+weak lensing model covering the core and the outskirts of the cluster. The first one is the necessity to include an external shear component to obtain a good model in the core. We have shown in Sect.~\ref{sec:subs_impact} and Sect.~\ref{sec:toy_models} that this external shear cannot be fully replaced by the presence of substructures in the cluster outskirts, and therefore that it introduces a non-physical aspect in our model. In addition, this external shear component creates an inconsistency between the parametric model in the core and the grid model in the outskirts. More generally, there can be some intrinsic difficulties in combining a parametric model and a grid. For instance, there could be some edge effects arising in the transition region between the fully parametric model in the core, and the weak-lensing grid. The impact of this effect is difficult to estimate precisely, but it might be responsible for positional uncertainties in the position of substructures that are located too close to the core (see Sect.~\ref{sect:sub_stars}).

To explore the impact of these two effects, we derive a weak-lensing-only and grid-only mass reconstruction of Abell\,370. We use the same grid structure as presented in Sect.~\ref{sec:grid_model} and shown in Fig.~\ref{fig:grid}, but without removing the grid points from the strong-lensing region. We superpose on this free-form model the galaxy-scale haloes corresponding to cluster galaxies, but do not optimize their parameters, fixing them instead to the best fit values obtained in Sect~\ref{sec:result_core}. We then optimize the amplitude of the grid potentials using the weak-lensing constraints. We refer to this model as model$_E$. This type of model has the advantage of not including an external shear component, and not having a transition between a parametric and free-form model. However, the lack of a parametric component that ``stabilizes'' the mass distribution in the core, and the lack of the high resolution strong-lensing constraints can lead to an increased noise in the modeling, and to a potential decrease of the signal and flattening of the profile in the core region.
As in previous models, we generate 1000 mass maps to estimate the statistical noise in the mass reconstruction. The contours corresponding to the mean mass map obtained are shown in red in Fig.~\ref{fig:comb_map}, along with the mass levels obtained from the Sequential-Fit model$_A$ (Sect.~\ref{sec:sequential_fit} blue) and Joint-Fit model$_{D1}$ (Sect.~\ref{sec:joint_fit}, green).

Most of the substructures are similarly detected in all three models, in particular S1, S2 and S5. We note that S2 and S5 are more extended in the grid-only model. The extended substructures, S6 and S7, are both present in the grid-only model, with still an extended mass distribution (i.e., not as peaky as for some of the other substructures). Conversely, S3 has a different shape in the grid-only model: instead of appearing as a separate substructure in the North of the cluster core, it takes the shape of an elongation of the core itself towards the North. This kind of elongation, connected to the core distribution can be more difficult to reproduce in a composite model, as they would be sensitive to the transition between the two model components. Even more drastically, S4 is not detected in the grid-only model. As this candidate substructure also has no optical or X-ray counterparts, this could indicate that it is a model artifact.

We also show in Fig.~\ref{fig:all_density_profiles} the azimuthally averaged surface mass density profile of this model, with the dot-dashed red line representing the mean model and its standard deviation with a red shaded area. The grid-only model presents a much lower amplitude at all scales than the other models, due to the lack of strong-lensing constraints and underlying large scale parametric components. 
As a comparison, we show on Fig.~\ref{fig:all_density_profiles} the average surface mass density profile measured with the Subaru weak-lensing data in \citet{umetsu2022} as a thin black solid line. Their model is consistent with our baseline Sequential- and Joint-Fit models in the inner region of the cluster, but presents a  significant deviation in the outer region. Starting at $\sim 800$\,kpc, their model is in much better agreement with our weak-lensing only, grid-only model. This suggests that either weak lensing only analyses consistently underestimate the mass profile in the cluster outskirts, or that our combined modeling overestimate the model in this region. We will examine this issue further in a follow-up study on simulated clusters.

    \subsection{Physical reality of the candidate substructures}
    \label{sec:physical_reality}

Throughout this paper, we have presented different measurements of the mass distribution in Abell\,370 (lensing maps, light distribution, X-ray gas distribution), and examined different lensing modeling techniques to check the consistency of the obtained mass maps. Here, we summarize what these different analyses can tell us about the physical reality of the substructure candidates identified in Sect.~\ref{sec:result_grid}, and discuss the implications it has on the cluster history and formation processes.

We start with S1, which is located to the East of the cluster core, at $\sim 860$\,kpc from its centre. This structure is present in all the different models (Sequential-Fit model$_A$ in Sect.~\ref{sec:sequential_fit}, parametric-only model$_{C1}$ in Sect.~\ref{sec:param_subs}, Joint-Fit model$_{D1}$ in Sect.~\ref{sec:joint_fit} and grid-only model$_E$ in Sect.~\ref{sec:grid_only}, see Fig.~\ref{fig:rgb_contours} and ~\ref{fig:comb_map}), but does not carry a significant amount of mass in the parametric-only model$_{C1}$ (see Table.~\ref{tab:subs_slwl}). In all the models that contain a grid component, it is very peaky, and is not connected to any elongated mass distribution: it is made of only one grid potential. In addition, this overdensity does not have a strong X-ray or luminous counterpart. All these arguments lead us to believe that this candidate is more likely a model artifact than a true physical substructure.  

The second candidate, S2, is located in the North-East, at a similar distance from the core, $\sim 820$\,kpc. This substructure is also detected in all the models, and also contains a more significant amount of mass in the parametric only model, $\Delta M \sim 1.6 \times 10^{13}\msun$. It is difficult to ascertain the presence of an X-ray counterpart for S2,  as it is located close to the foreground galaxy LEDA 175370 and two X-ray point sources (see Fig.~\ref{fig:xray_map}) that are contaminating the signal. 
In terms of optical counterparts, S2 is set between two peaks of the cluster light distribution (see Fig.~\ref{fig:rgb_contours}). It is therefore difficult to ascertain its physical reality, but we would be tempted to associate this detection with these light peaks. Notably, it can be seen in the BUFFALO image that it is located very close to a group of massive cluster members. 
As for the reason why the light and lensing peaks are offset, we see three possible explanations: (i) it could be due to the transition between the parametric and grid models, that could create some ``edge effects'' at the limit of the strong lensing region and shift the position of substructures located close to it; (ii) S2 is located between two light overdensities (North-East and South-West from the substructure, see Fig.~\ref{fig:rgb_contours}), and our model may have produced one ``effective'' substructure instead of two; and (iii) the detected substructure could correspond to a trail of dark matter tidally stripped from the infalling group, and which would be easier to detect because of the lower cluster background density at larger radii. 

The next substructure candidate, S3, is located in the North of the cluster core, at $\sim 1000$\,kpc from the centre. The stellar mass contained within 175\,kpc at its location represents a slight overdensity as compared to the stellar mass density of the cluster at this radius (see Fig.~\ref{fig:rho_star}), and it coincides with an elongated cluster light overdensity that extends from the cluster core to the North (Fig.~\ref{fig:rgb_contours}). In addition, the grid-only model also presents a mass overdensity in this region, and has the shape of an elongated structure connected to the core (Fig.~\ref{fig:comb_map}), similar to what appears in the lightmap. As discussed previously, this can be a limitation of the parametric+grid type of modeling, where it can be difficult to properly represent the transition between the two model regimes. We will investigate this in future work on simulated clusters. There is no X-ray counterpart at the location of S3, and any elongated structure between S3 and the cluster core would be masked by the presence of the bright foreground galaxy.

The candidate substructure S4 is located closer to the core, at $\sim 665$\,kpc in the South-West. It is detected only in the Sequential and Joint Fits, but not in the grid-only or parametric-only models (see Fig.~\ref{fig:rgb_contours} and \ref{fig:comb_map}). It has no X-ray counterpart, and no light peak located in its vicinity, which leads us to believe that this is also a model artifact.
 
Finally, substructure S5 is located in the south of the cluster core, at $\sim 890$\,kpc. It is strongly detected in all the models, and coincides with a stellar mass overdensity as compared to the cluster stellar mass density at this cluster-centric radius (Fig.~\ref{fig:rho_star}). There is however no counterpart detectable in the X-rays, and it is difficult to conclude if this overdensity corresponds to a true substructure, in the sense of a group of galaxies bound together, or if it corresponds to a projection of a few galaxies with halos massive enough to create some additional lensing boost. To disentangle that, it could be useful to obtain spectroscopic redshifts for the galaxies composing this overdensity. To that end we are currently proposing for a new, wide-area spectroscopic survey (called BUFFALO-WINGS) designed to cover the entire BUFFALO cluster area with MUSE spectroscopy. Like the outermost regions of the current data set, BUFFALO-WINGS would be shallow, but the exposure depth would be more than enough to confirm the (luminous) distribution of S5, as well as any other substructure candidate where we find possible cluster members.

We now discuss the two extended substructures. S6 is closest to the cluster core, at $\sim 630$\,kpc in the South/South-East direction, and is detected in all the models. Because of its extended nature, it is difficult to exactly pinpoint its centre, which we believe leads to the fact that its stellar mass content does not represent an overdensity as compared to the cluster density at that radius (Fig.~\ref{fig:rho_star}). When looking instead at the cluster lightmap, in Fig.~\ref{fig:rgb_contours}, it does present an overdensity to the East of the identified centre of S6, which coincides with the lensing detected extension. However, there is no cluster member galaxy overdensity at the exact location of S6. This is even more puzzling as there is an X-ray counterpart for this substructure. Its peak position overlaps with an extension in the cluster X-ray luminosity, which translates into an overdensity in the luminosity profile computed in this direction (Fig.~\ref{fig:xray_profiles}). More notably, there is a coincident region with low X-ray temperature and low entropy (see Fig.~\ref{fig:xray_maps}). These combined data sets suggest that S6 is a substructure infalling for the first time into the cluster, along a South to North trajectory. The temperature map could also suggest that there is a trail of low temperature stripped gas following in the South of S6, but this interpretation is more putative.  The lack of a clear counterpart in the galaxy distribution dictates that some further analyses are still necessary to fully understand this region of the cluster.
 
\begin{figure}
\begin{center}
\includegraphics[angle=0.0,width=\linewidth]{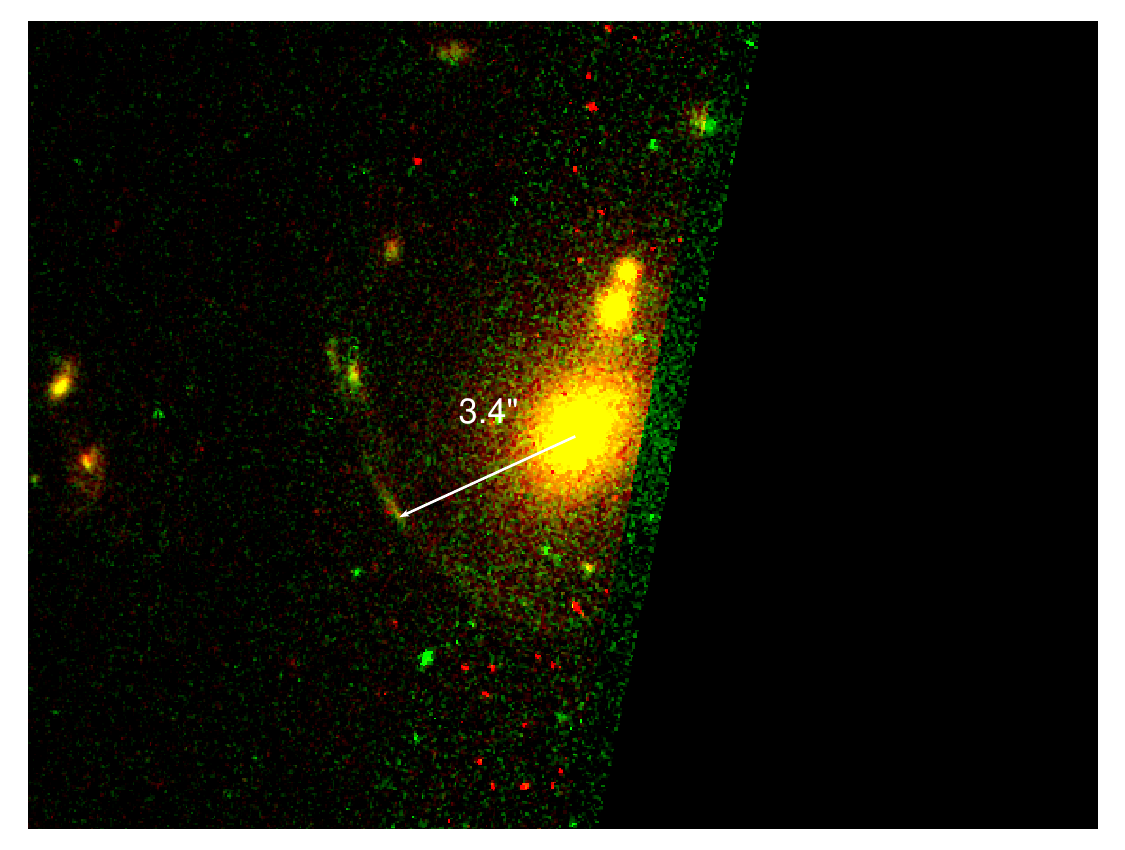}
\caption{Galaxy-galaxy strong lensing system appearing at the location of the substructure S7 \citepalias[system E in][]{lagattuta2022}. }
\label{fig:ggl_E}
\end{center}
\end{figure} 
 
S7, which is located at $\sim 780$\,kpc North-West from the cluster core, is similarly detected in all the models. It also matches with a stellar mass and light overdensity (Fig.~\ref{fig:rho_star} and ~\ref{fig:rgb_contours}), as well as with an extended X-ray luminosity overdensity (see  Fig.~\ref{fig:xray_map} and Fig.~\ref{fig:xray_profiles}). Given its extended nature, and the fact that it does not correspond to a low temperature or low entropy region, it would suggest that S7 is the remnant of an older merger, in which the gas has already been partly virialized in the cluster. 
In addition, there is a galaxy-galaxy strong lensing (GGSL) event which appears in this substructure, located at the edge of the BUFFALO field-of-view, but still visible in the BUFFALO imaging \citep[system E in ][]{lagattuta2022}. As shown on Fig.~\ref{fig:ggl_E}, the Einstein radius of this system is $\sim 3\farcs5$, which is significantly larger than what is expected by galaxy-galaxy lensing produced by a single galaxy (typically $\sim 1.5 - 2\arcsec$). Considering that the multiply imaged galaxy is located around redshift 1-2, this corresponds to a mass enclosed within the Einstein radius of $\sim 2 - 3\times 10^{12}\msun$. Assuming an NFW mass distribution in the lens, this corresponds to a mass of a few $10^{13}\msun$ enclosed within 175\,kpc, which is in very good agreement with our model.   The presence of this GGSL with a large Einstein radius therefore confirms that there is a matter overdensity at this location, which gives the extra-lensing boost responsible for this event. We note that the position of this substructure  is labelled at the centre of the extended mass overdensity rather than at the position of the GGSL system to be consistent with the other substructures.

In summary, combining the different modeling techniques presented throughout the paper, as well as the different observational probes, allows us to qualitatively assess whether each candidate substructure corresponds to a physical mass overdensity. After evaluating each substructure individually, it appears that the cluster presents an extension towards the North/North-West and South-East. Recently, \citet{ghosh2021} presented a non-parametric model for Abell\,370, optimized using the BUFFALO \emph{gold} strong-lensing constraints. As they did not include weak-lensing constraints, their model is focused on the core of the cluster, but it includes some mass clumps located $\gtrsim 200$\,kpc North and South of the core. They interpret these mass distributions as fictitious and generated by their model to account for some true mass distribution located outside of their modeling field-of-view. They show that their reconstructed mass distribution is consistent with a filamentary-like structure, extending North and South from the cluster core, in accordance with our model. 

Another recent model of Abell\,370 was presented in \citet{umetsu2022}, taking advantage of the wide field of the $BR_\mathrm{C}z'$ Subaru/Suprime-Cam weak-lensing data  over $\sim 30\arcmin\times 25\arcmin$.
In \citet{umetsu2022}, they reconstructed the cluster mass distribution in a free-form manner by combining weak-lensing shear and magnification constraints derived from the Subaru/Suprime-Cam data.  As the presented model is smoothed with a gaussian kernel of $1\farcm2$ it does not allow a direct comparison with the substructure detected in this work. However, some features are similar between the two reconstructions: their overall mass distribution presents boxy-shaped mass contours, with elongations towards the North-East/North-West and South-East/South-West, similar to the directions of our S2/S7 and S5/S6 substructures.

Finally, we note that it may seem surprising that only the two extended substructures show clear X-ray counterparts, as opposed to the five compact ones. We plan to run our lens model methods on mock clusters, in order to set up more quantitative criteria to differentiate true cluster substructures from model artifacts. It is possible that compactness of the candidates could be part of these criteria, meaning that for instance substructures that are detected as only one grid potential have more chance to be a model artifact. In any case, in the current state of our analysis, we cannot quantitatively ascertain whether S1, S2, S3, S4 and S5 are true cluster substructures. If they (or at least some of them) do correspond to physical mass overdensities, they would represent old structures that have been accreted into the cluster a long time ago. Indeed, their lack of X-ray counterparts (except maybe for S2) would suggest that they have spent enough time in the cluster for their gas to be completely virialized.

\subsection{What about the external shear?}
\label{sec:external_shear}

Having identified the most likely physical substructures in the cluster outskirts, we run one more lens model to quantify their impact on the mass reconstruction in the core. Similarly as for model$_{B1}$ in Sect.~\ref{sec:fixed_subs}, we re-optimize the parametric model in the cluster with the strong-lensing constraints only, and include the grid mass distribution as a fixed component of our model. In this case however, we remove all grid potentials which contribute to S1 and S4, keeping only the more reliable substructures, and call this model$_{B3}$. This configuration indeed allows to account for part of the external shear, reducing its value by half as compared to the baseline model, $\Gamma = 0.068^{+0.002}_{-0.004}$, with RMS=1.07\arcsec. Combining this result with all the models presented above and the toy models, we conclude that no physically reasonable substructures in the BUFFALO field-of-view could fully account for the model external shear. The orientation of the remaining shear field would require a mass distribution in the North-North-West or South-South-East. Interestingly the wide field mass model presented in \citet{umetsu2022} does show a massive overdensity at $\sim 1.5$\,Mpc in the North-West of the cluster, that could possibly further contribute to the shear. 

As described in Sect.~\ref{sec:xray_data}, we detect this structure in the \textit{XMM-Newton} X-ray observations, and measure its mass as $M_{500} = 1.2\times 10^{14}M_{\odot}$. We quantify the possible contribution to the external shear from this structure using the same method as in Sect.~\ref{sec:toy_models}, by generating 2D shear maps of the shear it produces and measuring the amplitude of this shear in the cluster centre. We model the structure as a NFW potential located at 7.4\arcmin\ from the cluster centre, and at a redshift $z = 0.32$, with mass $M_{500} = (1.2 \pm 0.3) \times 10^{14} M_{\odot}$. As we do not have constraints on the shape of this potential, we generate multiple shear maps, varying the concentration between $c_{500} = 3$ and 9. We find that due to its relatively large separation from the cluster core, this structure contributes only marginally to the total external shear, $\Gamma \sim 0.01$. However, we note that we only model this structure as one potential, while there could actually be an elongated mass distribution connecting the two structures, that could contribute further to the shear.

Alternatively, part of this external shear could be generated by some line-of-sight structures. This appears to be possible in MACS\,0717, which is less constrained by the strong lensing data \citep{williams2018}. However, in the case of Abell\,370, this scenario was thoroughly explored in \citet{lagattuta19}: although they identified a background structure composed of 35 galaxies at redshift $\sim 1$, and two smaller overdensities in the foreground, none of them allow to account for a significant part of the external shear component in any of their models.

Finally, rather than being the effect of some external structures (along the line-of-sight or outside of the strong lensing region), the external shear component could also compensate for the lack of flexibility of the parametric potentials, and act as a perturbation of their simple elliptical shapes. Alternative approaches have been explored to expand the flexibility of parametric mass distributions, such as including a free-form surface of B-spline functions that act as small perturbation of the parametric model \citep[][]{beauchesne2021}. This method has been applied to observed clusters in \citet{limousin2022}, and have allowed to improve their modeling without the use of an external shear component. It would be interesting to apply this method to Abell\,370 in the future and verify if it allows to eliminate the remaining shear.

\section{Summary and Conclusion}
\label{sec:conclusion}
In this paper, we have presented a strong+weak lensing analysis of the massive cluster Abell\,370, using data from the \emph{Beyond the Ultra-deep Frontier Fields And Legacy Observations} (BUFFALO) programme. BUFFALO is a treasury \textit{HST} programme that extends the spatial coverage with \textit{Hubble} of the 6 Frontier Field clusters by almost a factor of 4. These observations add deep weak-lensing data to the already existing high-precision strong-lensing constraints. Here we take advantage of these high quality data-sets to model the core and the outskirts of this massive and complex cluster. Our objectives are twofold: (i) extend the mass modeling to larger radii and detect possible substructures in the cluster's outskirts to better understand its formation/evolution history; and (ii) examine the impact of the outskirts mass distribution on the modelling of the cluster core. In particular, we aim to check whether modelling the cluster outskirts properly would reveal the physical origin of the external shear component that was introduced in \citetalias{lagattuta19}. We summarise our main findings here:
\begin{itemize}
    \item We have presented the different data products necessary for the mass modeling, obtained using the BUFFALO data-set. The strong-lensing catalogue is compiled from previously published strong-lensing candidates \citep[in][]{lagattuta17, lagattuta19, kawamata18, diego2018}, that were revoted and re-homogenized within the BUFFALO collaboration. The cluster members and weak-lensing catalogues are measured in the wide BUFFALO field-of-view, as described in Sect.~\ref{sec:cluster_gals} and \ref{sec:wl_constraints}, respectively. All our data products are available at MAST as a High Level Science Product\footnote{Data and lens models available via \url{https://doi.org/10.17909/t9-w6tj-wp63}}.
    \item Using these data products, we first construct and optimize our baseline mass model using the \textsc{Lenstool} software. It is a combination of two models, each based on a different approach for mass decomposition: (i) a parametric model in the core of the cluster, where the mass is decomposed into a small number of physically motivated mass components (see Sect.~\ref{sec:param_model} for more details). The position and shape of this part of the model is optimised with strong-lensing constraints. And (ii) a grid of mass ``pixels'' covering the cluster outskirts, adding flexibility to the mass distribution in this region, and optimised with the weak lensing constraints (Sect.~\ref{sec:grid_model}). We perform these two fits sequentially, meaning that we first model only the cluster core, then fix this component to its best fit, and optimise the amplitude of the grid mass pixels in the cluster outskirts in a second step. The resulting model, presented in Sect.~\ref{sec:sequential_fit}, has an external shear component in the cluster core, with an amplitude $\Gamma = 0.107$, and  five compact substructure candidates in the outskirts, as well as two more extended ones. Out of these seven substructures, four have a corresponding counterpart in the cluster light distribution, and two in the X-rays temperature and entropy maps. The case of substructure S6 is particularly interesting: there is mass extension to the East of the identified centre, that matches with an extended overdensity in the cluster member distribution. However, there is no cluster members detected at the position of the centre itself, while it corresponds to a clear signal in the X-ray data (see Appendix~\ref{sec:s6_xray}), which will require further analyses to be gully understood. For all the substructures, we combine the results for the different mass tracers considered (lensing, cluster member distributions and masses, X-rays) as well as alternative modeling approaches that we explore throughout the paper, and qualitatively assess the physical existence of each candidate substructures. We conclude that out of the seven, two may be model artifacts (Sect.~\ref{sec:physical_reality}). Considering only the more probable candidates, Abell\,370 appears to present some extended mass distribution towards the North/North-West and the South-East, in broad agreement with other recent mass reconstruction of the cluster \citep[][]{ghosh2021, umetsu2022}. 
    \item We explore the impact of this mass distribution measured in the cluster outskirts on the core of the cluster, and use three different approaches: (1) we fix the grid to its best fit, and re-optimize the core model taking into account these extra mass distributions (Sect.~\ref{sec:fixed_subs}); (2) we replace the grid model by five parametric potentials with very broad priors on their positions and velocity dispersions, to mimic the presence of substructures (Sect.~\ref{sec:param_subs}); and (3) we perform a combined model where the core and the outskirts are optimized jointly, using the \emph{hybrid}-\textsc{Lenstool} extension that we presented in \citet{niemiec2020} (Sect.~\ref{sec:joint_fit}). Although these alternative models do reduce the amplitude of the external shear in the cluster core, none of them remove it completely. A summary of all the different models discussed in the paper is presented in Appendix~\ref{sec:model_summary}. For all the models presented in this paper, the external shear component is oriented along the same direction, with $\theta_{\Gamma}$ in the range $17\degree - 20\degree$. To explain this shear by the presence of neighbouring structures, the model would require mass distributed along the North-West/South-East axis, which is consistent with what we found. However, the amount of mass located in these substructures would need to be much higher than what we detect to account for such a high value of external shear.
    \item We then examine if the shear included in the core may impact the detection and spatial distribution of the substructures in the cluster outskirts. For this, we construct a fully non parametric model, i.e. a grid covering both the core and the outskirts. We optimize this model using only the weak lensing constraints, and find a similar spatial mass distribution as in our baseline model. This suggests that some effects, such as the edge effects between the grid and the parametric model, do not impact the existence of most of the detected substructures, although they can add some uncertainties on their exact positions and shapes (Sect.~\ref{sec:grid_only}).
    \item In order to better understand the amount of substructure necessary to produce such a strong shear in the cluster core, we explore some toy models, with different spatial configurations and masses of substructures (Sect~\ref{sec:toy_models}). We found that it would require two aligned substructures, each with a mass $\sim 9 \times 10^{12} \msun$, located quite close to the core, $\sim 270$\,kpc, which seems an unlikely configuration for Abell\,370. In addition, the shear produced by these substructures would not be uniform over the whole modeled field, as is the case of the external shear parameter. This, together with the results outlined in the previous items, suggests that the external shear component cannot be fully accounted for by the presence of physically motivated substructures in the cluster outskirts. To replace it by a more physically motivated mass component may for instance require to use a more sophisticated mass distribution in the cluster core instead of a combination of elliptical potentials. One possibility would be to add small perturbations to increase the flexibility of the parametric mass distribution, as described in \citet{beauchesne2021,limousin2022}. 
\end{itemize}

In this work, we have presented multiple lens modeling approaches, in order to best assess the physical reality of the candidate substructures, but the final evaluation remains qualitative. Our next step would be to run some additional in-depths tests of the different modeling techniques presented here on simulated clusters, in order to develop more quantitative assessments, and conduct a full analysis of the systematic errors affecting the different types of modeling. In parallel, we are planing to apply comparatively the Joint-Fit and Sequential-Fit modeling methods on real, but slightly less complex clusters, such as Abell\,S1063 and MACS\,0616, in order to disentangle the different sources of modelling complexity.

\section*{Data Availability}
All the data products described and used in this work are released publicly with the paper, and are available at MAST as High Level Science Products via \url{https://doi.org/10.17909/t9-w6tj-wp63}. In addition, we provide at the same location outputs corresponding to the different mass models described in the paper.

\section*{Acknowledgements}
 AN and MJ are supported by the United Kingdom Research and Innovation (UKRI) Future Leaders Fellowship (FLF), `Using Cosmic Beasts to uncover the Nature of Dark Matter' (grant number MR/S017216/1). 
 KU acknowledges support from the Ministry of Science and Technology of Taiwan (grant MOST 109-2112-M-001-018-MY3) and from the Academia Sinica Investigator Award (grant AS-IA-107-M01). 
 AA has received funding from the European Union's Horizon 2020 research and innovation programme under the Marie Sklodowska-Curie grant agreement No 101024195 - ROSEAU. 
 J.M.D. acknowledges the support of projects PGC2018-101814-B-100 (MCIU/AEI/MINECO/FEDER, UE) Ministerio de Ciencia, Investigaci\'on y Universidades.  This project was funded by the Agencia Estatal de Investigaci\'on, and Unidad de Excelencia Mar\'ia de Maeztu, ref. MDM-2017-0765.
 DH acknowledges support by the ITP Delta foundation. 
 GM acknowledges funding from the European Union's Horizon 2020 research and innovation programme under the Marie Sklodowska-Curie grant agreement No MARACHAS - DLV-896778.
AZ acknowledges support by Grant No. 2020750 from the United States-Israel Binational Science Foundation (BSF) and Grant No. 2109066 from the United States National Science Foundation (NSF), and by the Ministry of Science \& Technology, Israel.
This project used the DiRAC Data Centric system at Durham University, operated by the
Institute for Computational Cosmology on behalf of the STFC DiRAC HPC
Facility (\url{www.dirac.ac.uk}). This equipment was funded by BIS
National E-infrastructure capital grant ST/K00042X/1, STFC capital
grant ST/H008519/1, and STFC DiRAC Operations grant ST/K003267/1 and
Durham University. DiRAC is part of the National E-Infrastructure.
This research is based on observations made with the NASA/ESA Hubble Space Telescope obtained from the Space Telescope Science Institute, which is operated by the Association of Universities for Research in Astronomy, Inc., under NASA contract NAS 5-26555. These observations are associated with program GO-15117.




\bibliographystyle{mnras}
\bibliography{reference} 


\appendix

\section{Summary of models}
\label{sec:model_summary}

Throughout the paper, we try different mass modeling approaches, meaning that we use different decompositions of the total mass distribution of the cluster, in order to test different sources of systematic errors. We summarize here the different components that are used at some point:
\begin{itemize}
    \item The "core parametric model": mass distribution in the central, strong-lensing region of the cluster described as a small number of haloes, i.e 4 cluster-scale haloes + 2 BCG haloes + 4 haloes corresponding to cluster member galaxies modelled individually.
    \item The "cluster member catalogue": all remaining cluster member galaxies, whose properties are modeled jointly with global scaling relations.
    \item The "non-parametric grid" of mass pixels, used to model the mass distribution in the cluster outskirts in a flexible manner.
    \item The "parametric substructures": instead of being modelled with the grid, the mass contained within the substructures in described with a small number of parametric haloes in some models.
\end{itemize}

\begin{table*}
    \centering
    \begin{tabular}{c | c c c c c c c c}
Model                       & ID & Optimisation  & $\Gamma$  & Section                   & RMS$_{\rm{SL}}$ ['']   & $\chi^2_{\rm{SL}}$ (dof)   & $\chi^2_{\rm{WL}}$ (dof)    & $\log(E)$ \\
\hline      
\multirow{2}{*}{Sequential} & A  & image         & 0.107     & \ref{sec:sequential_fit}  & 0.90          & 340 (102)     & 91333 (35558) & -34614    \\
                            &    & source        & 0.109     & \ref{sec:joint_fit}       & 1.56          & 437 (104)     & 91513 (35558) & -34709    \\
\hline
\multirow{2}{*}{Fixed subs} & B1 & image         & 0.096     & \ref{sec:fixed_subs}      & 0.98          & 400 (104)     &    --         &   --      \\
                            & B2 & image         & --        & \ref{sec:fixed_subs}      & 1.42          & 837 (106)     &    --         &    --     \\
Fixed subs, phys only       & B3 & image         & 0.068     & \ref{sec:external_shear}  & 1.07          & 477 (104)     &    --         &     --    \\
\hline
\multirow{2}{*}{SL+WL 
param only}                 & C1 & image         & 0.040     & \ref{sec:param_subs}      & 1.17          & 533 (37192)   & 91537 (37192) &   --      \\
                            & C2 & image         & --        & \ref{sec:param_subs}      & 1.19          & 832 (37200)   & 91487 (37200) &   --      \\
\hline
\multirow{2}{*}{Joint}      & D1 & source        & 0.09      & \ref{sec:joint_fit}       & 1.57          & 466 (35663)   & 91683 (35663) & -34719    \\
                            & D2 & source        & --        & \ref{sec:joint_fit}       & 1.84          & 243 (35665)   & 91381 (35665) & -34226    \\
\hline
Grid only                   & E  & --            & --        & \ref{sec:grid_only}       & --            &   --          &  --           & -34467    \\

    \end{tabular}
    \caption{Summary of the different models considered throughout the paper. The columns present, in order: the name of the model; its ID; the optimisation method for the parametric component (source vs image plane);  if any, the amplitude of the best-fit external shear component $\Gamma$; the section in which the model in discussed; the strong-lensing RMS value in arcseconds, the strong-lensing and weak-lensing $\chi^2$ with the degree of freedom (dof). If the model is optimised jointly on the weak- and strong-lensing constraints, the same dof value is given for both $\chi^2$. \textbf{Short IDs are attributed to each model to facilitate the matching with the publicly available output maps.}}
    \label{tab:model_summary}
\end{table*} 

These mass components are used in different combinations throughout the paper, and we give a summary of the different models in Table~\ref{tab:model_summary}. The different models are:
\begin{itemize}
    \item the Sequential Fit, which is the baseline model, where the parametric model in the core is first optimised with the strong-lensing constraints. It is then fixed to its best-fit parameter values, and the mass distribution in the cluster outskirts is the modelled in a second step with a non-parametric grid, constrained using the weak-lensing data. There are two realisations of this model, with the parametric model  optimised in the image plane and in the source plane. This model is composed of the "parametric core model" + the "cluster member catalogue" + the "non-parametric grid" in the outskirts.
    \item ``Fixed subs'': a re-optimisation of the parametric model describing the mass distribution in the cluster core, with the grid component in the cluster's outskirts fixed to the mass distribution obtained in the Sequential-Fit. Two versions are presented: with the external shear amplitude left to vary, and without the shear component. An additional run of the model is presented (``Fixed subs, phys only'') which contains only the mass contained in the substructures that are more likely to be true mass components, rather than model artifacts. This model contains the same components as the Sequential Fit, but with the "non-parametric grid" being fixed and not optimized.  
    \item ``SL+WL param only'': the mass distribution over the whole field is modelled with a parametric model, meaning that substructures in the outskirts are included as parametric potentials. This model was also optimised with and without the external shear component. This model contains the "parametric core model" + the "cluster member catalogue" + the "parametric substructures".
    \item The Joint Fit, where the parametric model in the core and the grid in the outskirts are optimised jointly, using both strong- and weak-lensing constraints. The parametric component of this model is optimised in the source plane, and the whole model was optimised with and without the external shear component. By definition this model is made of the same components as the Sequential Fit.
    \item ``Grid only'': a fully non-parametric model, optimised with the weak-lensing constraints only. It contains the "non-parametric grid" covering the whole cluster field (i.e cluster core + outskirts), as well as the cluster member catalogue.
\end{itemize}

In Table~\ref{tab:model_summary}, we provide different metrics that allow to quantify the goodness of fit the models. However, due to the different approaches of modelling that are used, not all metrics are defined in a meaningful way for all the models. The different ones that are considered are the root mean square separation between the model-predicted and observed positions of the multiple images for the strong-lensing constraints (RMS); the corresponding strong-lensing $\chi^2$ with the model degree of freedom (dof); the weak-lensing $\chi^2$ with the degree of freedom for parametric models; and the bayesian evidence for grid-based models. The degree of freedom is the difference between the number of constraints and the number of free parameters for each model, while the number of constraints for the strong lensing $\chi^2$ is equal to $N_{\rm{constraints}} = 2(N_{\rm{img}} - N_{\rm{src}})$, where $N_{\rm{img}}$ is the total number of multiple images and $N_{\rm{src}}$ is the number of corresponding background sources.

\section{Stellar-to-halo mass relation for the satellite galaxies}
\label{sec:shmr}

Cluster member galaxies are subject to interactions specific to this very dense environment, that will impact their properties, as well as the properties of the dark matter subhaloes in which they are embedded. In particular, the tidal forces of the host cluster strip part of the subhalo's dark matter, in an outside-in fashion \citep[see e.g.][]{niemiec2017, niemiec2020}. The impact of this tidal stripping will depend on the galaxy's time since infall and orbit within the cluster, which cause a radial segregation in the stellar-to-halo mass relation measured satellite galaxies \citep{niemiec2022}. Because of this effect, it may be wrong to consider a single scaling relation to model all the cluster galaxies together, as is done in this paper. To quantify the impact of this effect, we measure the stellar-to-halo mass relation for the galaxies located in the modeled field-of-view, and quantify how much it could vary within this field.

\begin{figure}
\begin{center}
\includegraphics[width=\linewidth,angle=0.0]{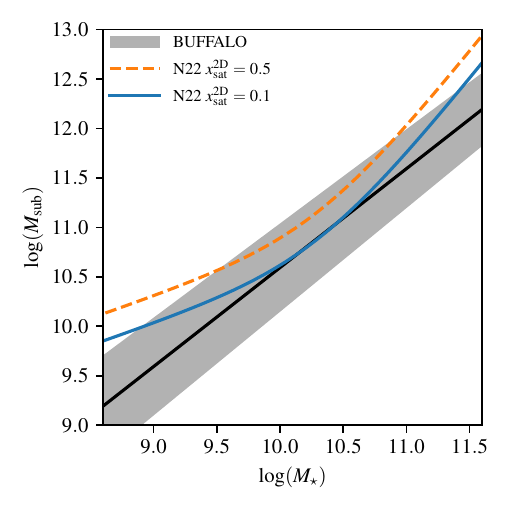}
\caption{Stellar-to-halo mass relation for the satellite galaxies in the BUFFALO field (black symbols). The stellar and subhalo masses are computed as described in the text. We show the error bars for only 1 in 10 points for clarity. For comparison, we show the stellar-to-halo mass relation calibrated on  the TNG simulation in \citet{niemiec2022}, for passive satellite galaxies, located at $x_{\rm{sat}}^{\rm{2D}} = R_{\rm{sat}}^{\rm{sat}}/R_{200} = 0.1$ (\emph{blue solid line}) and 0.5 (\emph{orange dashed line}).}
\label{fig:shmr}
\end{center}
\end{figure}

To compute the stellar-to-halo mass relation, we use the stellar mass estimates obtained from the F814W magnitude, as described in Sect.~\ref{sect:sub_stars}, and derive the corresponding subhalo masses from Equation 5 in \citet{jullo07}, using the best-fit values from the model in Sect.~\ref{sec:result_core} for $\sigma_0^{\star}$ and $r_{\rm{cut}}^{\star}$. The resulting relation for the satellites in the BUFFALO field-of-view are shown as the black line in Fig.~\ref{fig:shmr}. The error on this observed scaling relation is shown as a grey contour, and it accounts for the mean error on the galaxy magnitudes, as well as the uncertainty on the parameters describing the mass-to-light scaling relation, obtained from our mass model. As a comparison, we plot the stellar-to-halo mass relation calibrated on the TNG simulation in \citep{niemiec2022} for passive satellite galaxies, in a cluster at a redshift $z=0.24$. We show the relation computed at two different projected cluster-centric distances: $R_{\rm{sat}}^{\rm{2D}} = 0.1\times R_{200}$, which represent the very core of the cluster, and $R_{\rm{sat}}^{\rm{2D}} = 0.5\times R_{200}$, which is roughly the limit of the BUFFALO main field-of-view. Figure~\ref{fig:shmr} presents a good agreement between the observational measurement and the simulation result for the core galaxies. Even if this degree of agreement may seem surprising given the very different nature of the two measurements, and the number of approximations included in the observational estimation, it is what we would expect physically. Indeed, the strong-lensing constraints are much more sensitive than the weak lensing to the mass distribution in the subhaloes. It is therefore expected that the very stripped galaxies located close to the core will drive the fit of the scaling relation. The galaxies located further from the core will be less stripped, and follow a relation closer to the orange dashed line in Fig.~\ref{fig:shmr}. This relation is still relatively consistent with the observational measurement, given the error bars, but we still may be underestimating the mass of the subhaloes in the cluster outer regions. However, the flexibility of the grid model in the outskirts may still allow to include this ``missing'' mass. In future work, it will be interesting to include multiple scaling relations to account for tidal stripping, or even to implement a radially variable scaling relation, such as the one presented in \citet{niemiec2022}. We also note that the agreement between observations and simulations appears mostly in the intermediate mass range. This is due do the simplified estimation of the stellar mass from the F814W magnitudes, as well as the lack of flexibility in the scaling relation relating the subhalo mass to the galaxy magnitude. Further analyses would be necessary to obtain a more exact estimation of the BUFFALO satellite SHMR.

\section{X-ray properties of S6}
\label{sec:s6_xray}

As discussed in Sect. \ref{sec:xray_subs}, we detect a compact X-ray substructure at a position coinciding with the S6 lensing structure. Surprisingly, this structure does not have an obvious counterpart in the galaxy distribution and the accompanying X-ray structure is very compact, thus it is worth investigating the X-ray data around this position a little further. As already described in Sect. \ref{sec:xray_subs}, we detect a compact surface brightness excess at this position (see Fig. \ref{fig:xray_profiles}), and the thermodynamic map indicates the presence of a low-temperature (3-4 keV), low-entropy region at a position that is consistent with S6 (see Fig. \ref{fig:xray_maps}). However, given the resolution of \emph{XMM-Newton} it is in principle possible that the excess brightness observed at this position be due to a blend of point sources. 

To investigate whether one or several point sources in this area could have been missed, we searched the \emph{Chandra} archive for available high-resolution observations. Unfortunately, the existing ACIS-I observation of the cluster is very shallow. However, a relatively deep (88 ks) ACIS-S observation exists (observation ID 515). While the high background makes it difficult to use this observation to study the diffuse emission, the existing ACIS-S data are deep enough to detect any contaminating point source much below the flux limit of \emph{XMM-Newton}. No obvious point source is detected at the position of S6, such that we can conclude that the compact structure detected around S6 is not induced by one or several unmasked point sources.

\begin{figure}
\resizebox{\hsize}{!}{\includegraphics{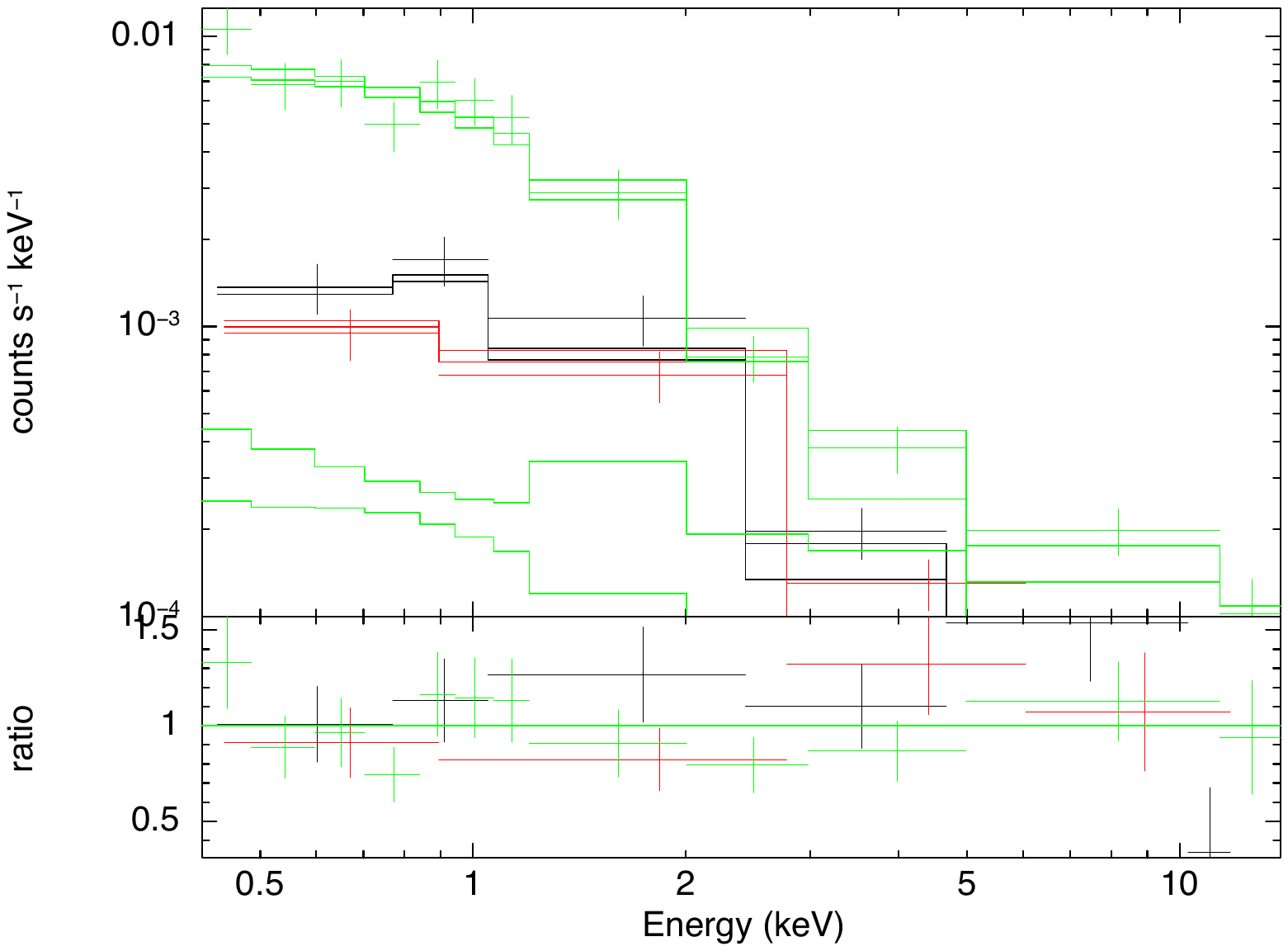}}
\caption{\emph{XMM-Newton} spectra of the region surrounding the S6 substructure. The data points indicate the spectra from the MOS1 (black), MOS2 (red), and PN (green) instruments, whereas the solid lines indicate the best fitting absorbed APEC model (see text). The bottom panel shows the residuals from the model.}
\label{fig:S6_xmm}
\end{figure}

As shown in Fig. \ref{fig:xray_maps}, the region surrounding S6 appears to have a low temperature ($3-4$ keV compared to $\sim10$ keV for the surrounding regions). While there is in principle sufficient signal to warrant that the temperature map in this region is accurate, we attempted to verify the temperature estimate in the region using a full-blown spectral modeling approach. To this aim, we extracted the spectrum of a region of 18$^{\prime\prime}$ radius around the position of S6 and fitted it using an absorbed APEC model (see Sect. \ref{sec:xray_data}). In Fig. \ref{fig:S6_xmm} we show the spectra of the region from the three EPIC instruments with the best fitting model superimposed. The best fitting model has a temperature of $3.99_{-0.81}^{+0.97}$ keV, which agrees with the value estimated in our temperature map. Therefore, we can conclude with confidence that the region surrounding S6 coincides with a region of enhanced X-ray surface brightness at a temperature that is significantly lower than the mean surrounding temperature ($\sim10$ keV). Such characteristics are consistent with a recent infall from S6 onto A370, with the core of the substructure not having mixed yet with the ambient ICM.

\bsp	
\label{lastpage}
\end{document}